\documentclass[preprint,12pt]{elsarticle}

\usepackage{amssymb}
\usepackage{amsmath}
\usepackage{amsfonts}
\usepackage{amsthm}
\usepackage{algorithm}
\usepackage{algpseudocode}
\usepackage{array}
\usepackage{bbding}
\usepackage{booktabs}
\usepackage[caption=false,font=normalsize,labelfont=sf,textfont=sf]{subfig}
\usepackage{textcomp}
\usepackage{stfloats}
\usepackage{float}
\usepackage{url}
\usepackage{verbatim}
\usepackage{adjustbox}
\usepackage{graphicx}
\usepackage{wrapfig}
\usepackage{multirow}
\usepackage{makecell}
\usepackage{xcolor}

\DeclareMathOperator*{\argmin}{arg\,min}
\newcolumntype{M}[1]{>{\centering\arraybackslash}m{#1}}

\newtheorem{theorem}{Theorem}
\newcommand{\bl}[1]{\textcolor{black}{#1}}

\journal{Neurocomputing}

\begin{document}

\begin{frontmatter}



\title{Optimized Layerwise Approximation for Efficient Private Inference on Fully Homomorphic Encryption}


%% Author name
\author[labelSNU]{Junghyun Lee\fnref{first}}
\author[labelC]{Joon-Woo Lee\fnref{first}}
%\ead{ljhfree530@snu.ac.kr}
\author[labelSejong]{Eunsang Lee}
%\ead{eslee3209@sejong.ac.kr}
\author[labelDG]{Young-Sik Kim}
%\ead{ysk@dgist.ac.kr}
\author[labelI]{Yongwoo Lee}
%\ead{yongwoo@inha.ac.kr}
\author[labelP]{Yongjune Kim\corref{cor}}
\ead{yongjune@postech.ac.kr}
\author[labelSNU]{Jong-Seon No}
%\ead{jsno@snu.ac.kr}

\fntext[first]{Equally contributed.}
\cortext[cor]{Corresponding author.}
\affiliation[labelSNU]{organization={Department of Electrical and Computer Engineering, INMC, Seoul National University},%Department and Organization
            city={Seoul},
            country={Korea}}

\affiliation[labelC]{organization={School of Computer Science and Engineering, Chung-Ang University},%Department and Organization
            city={Seoul},
            country={Korea}}

\affiliation[labelSejong]{organization={Department of Software, Sejong University},%Department and Organization
            city={Seoul},
            country={Korea}}

\affiliation[labelDG]{organization={Department of Electrical Engineering and Computer Science, DGIST},%Department and Organization
            city={Daegu},
            country={Korea}}

\affiliation[labelI]{organization={Department of Electrical and Electronic Engineering, Inha University},%Department and Organization
            city={Incheon},
            country={Korea}}

\affiliation[labelP]{organization={Department of Electrical Engineering, POSTECH},%Department and Organization
            city={Pohang},
            country={Korea}}

%% Abstract
\begin{abstract}
Recent studies have explored the deployment of privacy-preserving deep neural networks utilizing homomorphic encryption (HE), especially for private inference (PI). Many works have attempted the approximation-aware training (AAT) approach in PI, changing the activation functions of a model to low-degree polynomials that are easier to compute on HE by allowing model retraining. However, due to constraints in the training environment, it is often necessary to consider post-training approximation (PTA), using the pre-trained parameters of the existing plaintext model without retraining. 
Existing PTA studies have uniformly approximated the activation function in all layers to a high degree to mitigate accuracy loss from approximation, leading to significant time consumption. 
This study proposes an \emph{optimized layerwise approximation} (OLA), a systematic framework that optimizes both accuracy loss and time consumption by using different approximation polynomials for each layer in the PTA scenario. 
For efficient approximation, we reflect the layerwise impact on the classification accuracy by considering the actual input distribution of each activation function while constructing the optimization problem. 
Additionally, we provide a dynamic programming technique to solve the optimization problem and achieve the optimized layerwise degrees in polynomial time. 
As a result, we successfully approximated the ReLU and GELU functions, significantly reducing time latency while maintaining classification performance.
Especially, the OLA method reduces inference times for the ResNet-20 model and the ResNet-32 model by 3.02 times and 2.82 times, respectively, compared to prior state-of-the-art implementations employing uniform degree polynomials. 
%Furthermore, we successfully classified CIFAR-10 by replacing the GELU function in the ConvNeXt model with only 3-degree polynomials using the proposed method, without modifying the backbone model.
\end{abstract}

%% Keywords

\end{frontmatter}



\section{Introduction}
With the advancement of cloud computing, numerous data sets are being shared online, and data analysis is often carried out through data sharing via the cloud. One prominent example is Machine Learning as a Service (MLaaS), which involves performing machine learning tasks in the cloud.
However, sharing data via the cloud requires significant attention to privacy concerns.
Therefore, \textit{privacy-preserving machine learning} (PPML) is imperative to provide secure machine learning services while guaranteeing data privacy. One of the promising solutions in PPML is homomorphic encryption (HE). Clients can use this HE technique to encrypt their data, preventing the disclosure of personal information in the shared data on the cloud. By leveraging the properties of HE, servers can perform private inference (PI) using homomorphic operations directly on ciphertexts without decryption. Since fully homomorphic encryption (FHE) allows an unlimited number of operations on ciphertext, it is particularly well-suited for deep learning tasks requiring long sequential computations for many layers \cite{gilad2016cryptonets, ngraph, HCNN_GPU, strided, lee2023precise, kim2023optimized, pmlr-v202-lee23m}.

One of the important techniques in PPML using FHE is approximating non-arithmetic functions (e.g., ReLU, sigmoid, and GELU), since existing FHE schemes support only basic arithmetic operations such as addition and multiplication.
Such approximation techniques in PI can be categorized into two approaches: \emph{post-training approximation} (PTA), and \emph{approximation-aware training} (AAT). 
The PTA method focuses on utilizing existing well pre-trained networks by approximating activation functions with polynomials precisely, without modifying model parameters or structures \cite{lee2022privacy,strided,kim2023optimized}. The key is to approximate the given polynomial as accurately as possible, which has traditionally required the use of high-degree polynomials.
\bl{On the other hand, the AAT method aims to replace the activation functions of validated model with entirely different low-degree polynomials, such as only 2--3 degree polynomials, followed by retraining to update the parameters \cite{gilad2016cryptonets,FasterCryptoNets,HCNN_GPU,ali2020polynomial}.}
This approach \bl{requires} additional training computations or supplementary data to apply lower-degree polynomials. 

\bl{Although several experimental results have shown that AAT approaches can effectively reduce latency without compromising accuracy, there are two fundamental limitations. First, they require direct access to raw training datasets for fine-tuning. In practice, one may wish to perform inference on encrypted data using pre-trained and validated models. In such cases, retraining may be infeasible due to the lack of access to the original training data.
Second, the modification of activation functions becomes increasingly sensitive to performance as the dataset grows larger and more complex. Replacing activation functions with low-degree polynomial alternatives (as in AAT) is generally not effective for large datasets. The shape of activation functions is known to have a major impact on model performance, and there has been extensive research on fine-tuning activation shapes for accuracy. While AAT approaches often demonstrate effectiveness on small datasets, they fail to achieve competitive performance on large datasets, regardless of retraining, due to the intrinsic limitations of low-degree polynomial activations.
For these reasons, AAT research faces fundamental limitations at scale, whereas PTA research—which focuses on accurately approximating proven activation functions—offers a promising direction for designing privacy-preserving models that maintain high performance even on large datasets.}

% The two approximation methods cannot be compared as one being superior to the other, as each has its own clear advantages and disadvantages, making them both important PI approaches that should be studied independently.

This research aims to overcome the main bottleneck in the PTA approach: significantly prolonged inference time. Despite the advantage of avoiding additional computational overhead during model retraining, the PTA still suffers from large inference latency due to the use of high-degree polynomials.
The state-of-the-art PTA work \cite{strided} utilized a composite minimax polynomial approximation technique for the sign function \cite{leeminimax} to accurately approximate ReLU and successfully implemented ResNet models on the RNS-CKKS scheme. But, it replaces a single ReLU function with a polynomial of degree 6,075 (composition of polynomials with degree 15, 15 and 27), which requires a considerable amount of bootstrapping time and inference latency.

Prior PTA approaches determined the approximate polynomial without adequately considering the characteristics of each layer in deep neural networks. They subsequently employed the identical polynomial across all layers, rendering these implementations less efficient.
Considering the limitations of prior PTA works, we introduce a new principled approximation method for PTA in PI: the \textbf{Optimized Layerwise Approximation (OLA)} method. Unlike the previous PTA approach, which utilized the same polynomial for the activation functions of all layers, the proposed OLA method considers using approximation polynomials with varying degrees.  
Here, we incorporate the actual input distribution of each layer's activation function to reflect the differing impacts for each layer.
Considering those impacts, we formulate the optimization problem aimed at reducing inference latency without degrading classification accuracy.

\bl{According to our experiment results, the OLA method reduces the inference latency of the state-of-the-art method of 3.02 times and 2.82 times on ResNet-20 and 32, respectively. 
Additionally, the simulation results demonstrate that the GELU function in ConvNeXt can be effectively approximated using only degree 3 polynomials. 
By successfully replacing activations with only polynomials degree 3 without additional modification of the pre-trained network with non-arithmetic activations, the OLA method is expected to play a significant role in future PTA approach for PI.}

This paper is organized as follows. First, we analyze how each layer of the deep neural network contributes to overall classification accuracy and present a suitable optimization problem related to the polynomial degree in Section \ref{sec:weighted}. Second, we introduce a dynamic programming technique in Section \ref{chap:optprob}, to find the optimal degree solution of the optimization problem in polynomial time. Then, we describe the OLA method, how to apply the obtained optimal degree solution in Section \ref{sec:OLA}. Finally, we summarize the results of simulations, applying the obtained polynomials through the OLA method to ResNet and ConvNeXt in Section \ref{sec:result}.

\section{Related Works}
\bl{Our work falls under the research direction of privacy-preserving machine learning (PPML). Within PPML, a major line of research focuses on enabling efficient private inference through untrusted third parties while preserving data privacy during the inference phase. This research is typically pursued along three main directions: HE-based PPML \cite{ju2024neujeans,kim2023optimized,lee2023precise,strided,zhang2024secure,moon2024thor,park2024powerformer,ran2023spencnn,kim2023hyphen}, MPC-based PPML \cite{luo2024secformer,li2022mpcformer,maeng2024accelerating,kei2025shaft,xia2025cryptpeft,yuan2024md,cheng2025mosformer}, and hybrid PPML \cite{park2022aespa,huang2022cheetah,xu2024Privcirnet,pang2024bolt,hao2022iron,xu2025breaking,mishra2020delphi} that combines both techniques. Across all three directions, much effort has been devoted to efficiently handling or replacing non-arithmetic functions.}

\bl{At a high level, our work aims to reduce latency by effectively approximating non-arithmetic activation functions in situations where retraining is infeasible, building on encryption schemes that support only arithmetic operations. Therefore, our method is not limited to HE-based approaches, but can also be extended to MPC-based or hybrid PPML. Nevertheless, in this paper we restrict our focus to HE-based PPML, while noting that applying our technique to the other directions would be an interesting avenue for further work.}

\bl{HE-based PPML can be divided into two approaches: post-training approximation (PTA) \cite{ju2024neujeans,kim2023optimized,lee2023precise,strided,zhang2024secure,moon2024thor} and approximation-aware training (AAT) \cite{park2024powerformer,ran2023spencnn,kim2023hyphen}. AAT research has explored replacing non-arithmetic operations with HE-friendly alternatives via fine-tuning. Such operations include activation functions, normalization layers, softmax, and others. While various fine-tuning techniques have successfully replaced these blocks without accuracy degradation, activation functions have only been validated on simple datasets, and thus far have not demonstrated effectiveness on large datasets or complex models. This limitation reflects findings in mainstream model design research, where fine adjustments to activation shapes are known to significantly impact performance. Hence, although AAT can be effective for simple tasks, it faces fundamental limitations for more complex models, especially with respect to activation functions \cite{park2024powerformer}. Since our research directly addresses the efficient computation of activation functions, we believe the PTA direction is a more appropriate long-term approach.}

\bl{Within PTA, there are two major issues: (1) developing efficient homomorphic methods for linear operations such as convolution and matrix–vector multiplication \cite{ju2024neujeans,kim2023optimized,moon2024thor}, and (2) developing efficient methods for non-arithmetic operations such as ReLU and softmax \cite{strided,zhang2024secure,cho2024fast,park2024powerformer}. These two issues are orthogonal and complementary, since they target different blocks of computation, and thus each should be compared against techniques from its respective research line.}

\bl{For non-arithmetic operations, there are two primary approaches: using BFV/CKKS HE schemes \cite{meftah2021doren,strided,kim2023optimized}, which support arithmetic operations and approximate non-arithmetic functions with small error; or using TFHE \cite{lou2019she}, which supports bit-wise operations for exact computation. While TFHE avoids approximation error, it suffers from large memory overhead and poor compatibility with ciphertext packing, making it impractical on real-world servers. CKKS, by contrast, allows many values to be encrypted in a single ciphertext, providing strong advantages in runtime and memory efficiency, but it requires approximation of non-arithmetic functions. Our work addresses this limitation by adaptively adjusting approximation degrees per layer to balance classification accuracy and latency, thereby improving the efficiency of CKKS-based PPML.}

\bl{From this perspective, the most directly comparable work to ours is that of Lee et al. \cite{lee2023precise}. In particular, Lee et al. \cite{lee2023precise} showed that applying the minimax composition method from Lee et al. \cite{strided} to approximate ReLU in ResNet models can work in a PTA setting without retraining. However, their approach did not consider per-layer characteristics and instead applied a uniform approximation. Subsequent PTA works such as Lee et al. \cite{strided} and Kim et al. \cite{kim2023optimized} adopted this same uniform implementation while focusing on optimizing linear operations. Thus, the most recent PTA studies that explicitly address non-arithmetic operation optimization are Lee et al. \cite{lee2023precise} and its follow-ups, and accordingly we compare our work directly against them. Recently, there has been growing research on implementing transformer models with homomorphic encryption, and these studies also adopt a uniform approximation approach across layers. Since transformers contain many non-arithmetic operations in each layer, applying our technique to transformer models represents an important direction for further work.}

\section{Layerwise Degree Optimization Problem}
\label{sec:weighted}

We aim to adjust the approximate polynomials for each layer by reflecting how the degree of activation approximation in each layer affects the accuracy of the neural network. In this section, we characterize these effects and formulate an optimization problem with mathematical modeling. Our optimization problem is constructed by three steps: (i) weighted least squares approximation considering the distributions of each layer, (ii) estimation of the impact of each approximation on the loss function with the variance of the loss, and (iii) formulation of an optimization problem for the relationship between average loss variance and inference runtime.

\subsection{General Distribution-Aware Approximation}

Unlike the prior PTA approaches which involve the minimax approximation (considering $\ell_\infty$-norm), OLA method employs the weighted least squares approach (based on the $\ell_2$-norm), which better reflects the actual input distribution of each activation function.

\bl{Recently, AutoReP \cite{iccv} has \bl{replaced} the ReLU function with a degree-two polynomial using the weighted least squares method, taking into account the input distribution within the AAT approach.
% However, AutoReP uses the \emph{fixed} normal distribution $\mathcal{N}(0, 2)$ for \emph{every layer}. 
% In our observation, more efficient approximation is possible when we adopt the real input distribution in each layer rather than simply using an oversimplified distribution like $\mathcal{N}(0,2)$ (refer to Table \ref{tab:main_result} for this observation).
We apply the weighted least squares method in the PTA approach.
In contrast to the AAT approach, more accurate approximations in PTA may necessitate approximate polynomials with degrees greater than two.
Although AutoReP shows the coefficients of the second-order approximate polynomials, we generalize the closed form of the approximate polynomials with polynomial degree $d$ in the following theorem.}

\begin{theorem}
\label{thm:normal}
Let $\phi(x)$ be an arbitrary input distribution, and $\mathcal{F}_\phi$ be a function space $\{ g| \int_\mathbb{R} \phi (x) |g(x)|^2 dx < \infty \}$. Assume that every polynomial and a function $f(x)$ are elements of $\mathcal{F}_\phi$.
\begin{enumerate}[(a)]
    \item There exist polynomials $\{h_l(x)\}_{l\geq 0}$ where the degree of $h_l(x)$ is $l$ for all $l\geq 0$ and  
    $\int_\mathbb{R} \phi(x)h_m(x)h_n(x)dx=\delta_{mn}$
    for all $m,n\geq 0$, that can be computed in a polynomial-time algorithm.
    \item The approximate polynomial for the function $f(x)$,
\[
P_\phi[d;f](x):=\sum_{l=0}^d h_l \left( x \right) \int_\mathbb{R} \phi(t) f(t) h_l \left( t \right) dt,
\]
minimizes the MSE over the polynomials with a degree not greater than $d$, where $h_l(x)$ satisfies the property in (a).
\item The minimized MSE from (b), $E_{\phi}[d;f]$, is given by
\[
E_{\phi}[d;f] =  \int_\mathbb{R} \phi(x)f(x)^2dx 
- \sum_{l=0}^d \left[ \int_\mathbb{R} \phi(x)f(x)h_l \left( x\right) dx  \right]^2.
\]
\end{enumerate}
\end{theorem}
\begin{proof}
    For convenience, we write $\langle g_1, g_2\rangle_\phi := \int_\mathbb{R} \phi(x)g_1(x)g_2(x)dx$ for $g_1,g_2\in \mathcal{F}_\phi$. Then, $\mathcal{F}_\phi$ can be considered as a Hilbert space with inner product $\langle \cdot,\cdot \rangle_\phi$, which is a well-known fact in mathematical analysis. Also, we write $\|g\|_\phi := \langle g,g \rangle_\phi^{1/2}$ for $g \in \mathcal{F}_\phi$.
    
    (a) Considering $\mathcal{F}_\phi$ as a Hilbert space, we can perform Gram-Schmidt process from linearly independent set $\{1,x,x^2,\cdots\}\subseteq \mathcal{F}_\phi$. Then, the result of the process $\{h_0(x),h_1(x),h_2(x),\cdots \}\subseteq \mathcal{F}_\phi$ is an orthonormal subset of $\mathcal{F}_\phi$, which leads $\langle h_m, h_n\rangle_\phi=\delta_{mn}$. By the property of the Gram-Schmidt process, the process requires polynomial-time, and the degree of $h_l(x)$ is $l$.

    (b,c) For given integer $d$, an arbitrary polynomial $p(x)$ with a degree not greater than $d$ can be expressed as $p(x)=\sum_{l=0}^d b_l h_l(x)$ for some $b_l\in\mathbb{R}$. Approximating the function $f(x)$ as the polynomial $p(x)$, the MSE of $p(x)$ can be expressed as $\int_\mathbb{R} \phi(x) (f(x)-p(x))^2 dx = \| f-p\|_\phi^2$. Some algebra shows that
    \begin{align*}
    \| f-p\|_\phi^2=\langle f-p,f-p \rangle_\phi &= \langle f,f \rangle_\phi - 2\sum_{l=0}^d b_l \langle f,h_l\rangle_\phi + \sum_{l,l'=0}^d b_lb_{l'} \langle h_l,h_{l'} \rangle_\phi \\
    &= \left[ \|f \|_\phi^2 - \sum_{l=0}^d \langle f,h_l\rangle_\phi^2 \right] + \sum_{l=0}^d \left( b_l-\langle f,h_l\rangle_\phi \right)^2.
    \end{align*}
    Therefore, the MSE $\| f-p\|_\phi^2$ is minimized when $b_l=\langle f,h_l\rangle_\phi $ for all $l=0,\cdots,d$. As a result, the polynomial $\sum_{l=0}^d b_l h_l (x) = \sum_{l=0}^d h_l(x) \langle f,h_l\rangle_\phi = P_\phi[d;f](x)$ minimizes the MSE over the polynomials with a degree not greater than $d$, and the minimized MSE is $\|f \|_\phi^2 - \sum_{l=0}^d \langle f,h_l\rangle_\phi^2=E_\phi[d;f]$.
\end{proof}

We use the polynomial $P_{\phi}[d;f](x)$ as a approximate polynomial instead of non-arithmetic activation function $f$. In practical situation, we approximate the input distribution $\phi$ as a normal distribution, because the input distribution for each activation function usually follows the normal distribution. 
Additionally, we formulate an optimization problem to determine the most effective polynomial degrees based on the minimized MSE $E_{\phi}[d;f]$ derived in the theorem above. The next subsection describes the details.

\subsection{Average Loss Variance}
\label{subsec:loss}

When training a neural network, we perform backpropagation to compute gradients to update each weight and understand the impact of each weight on the loss function. Likewise, we devise a method to compute the impact of the activation approximation on the loss function using gradients obtained via backpropagation in this situation. However, there is an inherent difference between these two situations. If we establish the approximating polynomial for the activation function, the sign and magnitude of the deviation from the real value may fluctuate based on the input provided to the activation function. Therefore, it is difficult to adjust the approximation noise accurately by adjusting the approximating polynomial.
To assess how the approximation error in each layer affects classification accuracy, we should use another measure instead of the gradient of the loss function. We propose to use the \emph{variance of the loss} as a new target function in our optimization problem.

If the activation function is substituted with approximate polynomials for the encrypted data, it introduces approximation errors that alter the optimized loss. We now assume that the pre-trained model under plaintext has successfully minimized the loss, and thus we can regard the change in the loss function as a degradation of the classification accuracy. We use a relaxation whereby the approximation error of each activation function is treated as a random variable, although it is deterministic for a given input value. This assumption enables us to quantify the relation between the classification accuracy and the approximation error of each activation function. The justification for this relaxation lies in the consideration that the approximation error can be regarded as a random variable across different input data of all layers.      

Consider a pre-trained model with a loss function $\mathcal{L}$. This loss function depends on the activation nodes and can be expressed as $\mathcal{L}(\{a_{i,j}\})$, where $a_{i,j}$ denotes the $j$th activation node in the $i$th layer ($i=1,\cdots,N_L$, $j=1,\cdots,n_i$). When we replace the activation function with approximate polynomials, the activation node $a_{i,j}$ is changed into $a_{i,j}+\Delta a_{i,j}$, where $\Delta a_{i,j}$ is a random variable representing an approximation error. These approximation errors $\{\Delta a_{i,j} \}$ introduce noise to $\mathcal{L}$, which we denote as $\Delta \mathcal{L}:=\mathcal{L}(\{a_{i,j}+\Delta a_{i,j}\}) -\mathcal{L} (\{a_{i,j}\})$. By the first-order Taylor approximation, the noise in the loss function is given by
$\Delta \mathcal{L} = \sum_{i,j}\frac{\partial\mathcal{L}}{\partial a_{i,j}}\Delta a_{i,j}$.

Suppose that we employ an identical approximate polynomial for each layer. For a given layer $i$, each individual approximation error $\Delta a_{i,j}$ is identically distributed since the nodes $\{a_{i,j}+\Delta a_{i,j}\}$ represent the output of the identical $i$th approximate polynomial. 
Note that the variance of $\Delta a_{i,j}$ precisely aligns with the minimized $E_{\phi_i}[d_i;f]$ as defined above in Theorem \ref{thm:normal}. Here, $d_i$ denotes the degree of the $i$th approximate polynomial, and $\phi_i$ denotes the input data distribution. This fact comes from
\[
\text{Var}[\Delta a_{i,j}] = \mathbb{E}[\Delta a_{i,j}^2] - \mathbb{E}([\Delta a_{i,j}])^2 = E_{\phi_i}[d_i;f] - (\mathbb{E}[\Delta a_{i,j}])^2,
\]
and $\mathbb{E}[\Delta a_{i,j}] =0$.\footnote{It can be derived from $\int_\mathbb{R}\phi_i(x)(f(x)-P_{\phi_i}[d;f](x))dx=\langle f-P_{\phi_i}[d;f](x), 1\rangle_{\phi_i}$. Considering $h_0(x)\equiv 1$ and $\langle h_l,h_0 \rangle_{\phi_i}=\delta_{l0}$, $\langle P_{\phi_i}[d;f](x), 1\rangle_{\phi_i}=\langle f, 1 \rangle_{\phi_i}$.}
Furthermore, we assume that each random variable $\Delta a_{i,j}$ is independent. This assumption is reasonable given that each $\Delta a_{i,j}$ originates from independent polynomial approximations for each $i$th activation function, and each $a_{i, j}$ is computed independently using kernels for each $j$th activation node. Then, the variance of the loss function can be represented by
\begin{align} \label{eq:var_loss}
    \text{Var} [\Delta\mathcal{L}] = \text{Var} \left[  \sum_{i,j}\frac{\partial\mathcal{L}}{\partial a_{i,j}}\Delta a_{i,j}\right] \nonumber
    = \sum_{i,j}\left(\frac{\partial\mathcal{L}}{\partial a_{i,j}}\right)^2 \text{Var}[\Delta a_{i,j}] \nonumber = \sum_{i}\alpha_iE_{\phi_i}[d_i;f],
\end{align}
where $\alpha_i = \sum_j \left(\frac{\partial\mathcal{L}}{\partial a_{i,j}}\right)^2$ quantifies the impact of the $i$th layer's approximation error on classification accuracy. 
Note that the values of $\alpha_i$  are different for each image. If we denote $\alpha_{i, k}$ as the value for the $k$th image in training set with $N_T$ images, then we can calculate the average of $\alpha_{i,k}$'s for all images in the training set, denoted as $A_i=\frac{1}{N_T}\sum_{k} \alpha_{i,k}$. Our objective is to minimize the average loss variance function $V(\mathbf{d})=\text{Var} [\Delta\mathcal{L}] =\sum_{i=1}^{N_L}A_iE_{\phi_i}[d_i;f]$ by optimizing the set of degrees $\mathbf{d}=(d_1,\cdots,d_{N_L})$, where $N_L$ is the number of layers.

\subsection{Layerwise Optimization Problem with Average Loss Variance and Runtime}
\label{subsec:optproblemset}

We now consider the runtime for the \textit{ciphertext evaluation} of each layer when we use the set of approximate polynomial degrees $\mathbf{d}$. Assume that other parameters other than the degree set $\mathbf{d}$ are fixed, and if the activation function in the $i$th layer is approximated by $P_{\phi}[d;f](x)$ with degree $d$ then the runtime of this layer is denoted as $T_i(d)$. If we use the degree set $\mathbf{d}=(d_1,\cdots,d_{N_L})$, then the total runtime of the target neural network is $T(\mathbf{d})=\sum_{i=1}^{N_L} T_i(d_i)$. This runtime function $T_i$ for each layer can be set according to the neural network model type, the model parameters, and the implementation method with the homomorphic encryption scheme. 

It is sufficient to reflect only the factors in $T_i$ that may vary with the degree $d_i$. In the implementation of CNN in the RNS-CKKS scheme we used, the polynomial evaluation time and the bootstrapping time performed before the polynomial evaluation are major factors that depend on the polynomial degree $d$. The time required for other linear operations, such as convolution, is determined independently of changes in the degree of the polynomial. Therefore, we define $T_i(\cdot)$ as the sum of the polynomial evaluation time and the bootstrapping time for $i$th layer. Also, the first layer does not perform bootstrapping in our implementation, we define $T_1(\cdot)$ as the evaluation time for the first polynomial. See \ref{app:prelim} which describes the preliminaries about the bootstrapping of the RNS-CKKS scheme and the detailed implementation of CNN for the ciphertexts.

We now aim to formulate a layerwise degree optimization problem to co-optimize the average loss variance function $V(\mathbf{d})$ and the runtime function $T(\mathbf{d})$ with adequately setting the degree set $\mathbf{d}$ as follows: 
\begin{equation}
    \label{eq:opt_prob}
    \min_{d_1,\cdots, d_{N_L}\in\mathcal{S}}  \:  V(\mathbf{d}) = \sum_{i=1}^{N_L} A_i E_i(d_i)\quad \text{subject to} \quad T(\mathbf{d})=\sum_{i=1}^{N_L} T_{i}(d_i)\leq K,
\end{equation}
where $K$ is a constraint on inference runtime, and $\mathcal{S}$ is the degree search space. For simplicity, we represent $E_{\phi_i}[\cdot;f]$, the MSE of $f$ in Theorem 
\ref{thm:normal}, as $E_i(\cdot)$. 
Other factors besides the degree set $\mathbf{d}$, such as the mean and the variance of the input distribution, will be reflected when setting the optimization problem and fixed as constant when solving the problem. Note that $T_i(d_i)$ is an increasing function of $d_i$, as it takes more time to compute the larger degree polynomials. This fact will be used in solving this problem. 

\section{Dynamic Programming for Layerwise Degree Optimization}
\label{chap:optprob}

\subsection{Discretization of Constraint Function}
\label{subsec:disc}
Although the runtime function for the $i$th layer is simplified as a single-valued function on $d_i$ in the previous section, it is still difficult to represent the runtime function $T_i(d_i)$ as a closed form, so we should devise a numerical method for this optimization. However, the problem is still hard, since it can be regarded as a kind of discrete non-convex optimization. Instead of solving the exact optimization problem, we apply relaxation with discretization of the constraint function $T_i(d)$. It means that $T_i(d_i)$ is converted as $\tilde{T}_{i}^{\nu}(d_i):= \nu\lfloor T_{i}(d_i)/ \nu \rceil$, where $\nu$ is the discretization unit, and $\lfloor \cdot \rceil$ is the rounding function. Then, we now assume that $T_i(d_i)$ can be regarded as an integer multiple of the discretization unit $\nu$. We define $\tau^\nu_i (d_i) := \tilde{T}_{i}^{\nu}(d_i)/\nu = \lfloor T_{i}(d_i)/ \nu \rceil$ as the scaled integer of $T_i(d_i)$ relative to $\nu$. In other words, we have $\tilde{T}_{i}^{\nu}(d_i)=\nu\cdot \tau^\nu_i (d_i)$ and $\tau^\nu_i(d_i)\in \mathbb{N}$. When we do not focus on the value of the $\nu$ in $\tau_i^\nu(d_i)$, we can omit $\nu$ so that we can denote it as $\tau_i(d_i)$. 

With this relaxation, we tackle the optimization problem in \eqref{eq:opt_prob} via a \emph{polynomial-time} dynamic programming approach. Let's define $N_K := \lfloor K/\nu \rfloor \in \mathbb{N}$ and $\tau(\mathbf{d}) := \sum_{i=1}^{N_L}\tau^\nu_i(d_i)$. Then the optimization problem (\ref{eq:opt_prob}) can be relaxed to the following:

\begin{align}
    \label{eq:opt_prob_relax}
    \min_{d_1,\cdots, d_{N_L}\in\mathcal{S}}  \:  V(\mathbf{d}) = \sum_{i=1}^{N_L} A_i E_i(d_i)\quad \text{subject to} \quad \tau(\mathbf{d})=\sum_{i=1}^{N_L} \tau_{i}(d_i)\leq N_K,
\end{align}

where $\tau_i(d_i)$'s and $N_K$ are all positive integers. We will now solve \eqref{eq:opt_prob_relax} instead of the original optimization problem.
\subsection{Relation Between Subproblems}
To design the dynamic programming algorithm for the optimization problem \eqref{eq:opt_prob_relax}, we should set the recurrence relation between the subproblems. Suppose that we want to solve \eqref{eq:opt_prob_relax}, and we denote this problem as $\mathcal{P}(N_L, N_K)$. To be more general, the optimization problem $\mathcal{P}(l, k)$ can be defined as follows: 
\begin{align}
    \min_{d_1,\cdots, d_{l}\in\mathcal{S}}  \:  V(\mathbf{d}) = \sum_{i=1}^{l} A_i E_i(d_i)\quad \nonumber \text{subject to} \quad \tau(\mathbf{d})=\sum_{i=1}^{l} \tau_{i}(d_i)\leq k.
\end{align}
Then, we now consider the relation between $\mathcal{P}(N_L, N_K)$ and the subproblems $\{\mathcal{P}(l, k)\}$ with $l = 1,\cdots,N_L - 1$ and $k = 1, \cdots, N_K $. Let  $\mathbf{D}(l,k)=(D_1(l,k),\cdots,D_l(l,k))$ represent the optimal solution for $\mathcal{P}(l,k)$, where $D_i(l,k)$ denotes the degree of the $i$th polynomial for $i =1,2,\cdots,l$. Then, our main objective is to determine the optimal $\mathbf{D}(N_L,N_K)$, i.e., the solution of \eqref{eq:opt_prob_relax}.
In some cases, there may be no degree pairs $d_1,\cdots,d_l$ that satisfy $\sum_{i=1}^l \tau_i(d_i) \leq k$. To account for this situation, we set $\tau_i(-1):=0$, $E_i(-1):=\infty$ and include $-1$ in $\mathcal{S}$.

Let's induce the relation between $\mathbf{D}(l,\cdot)$ and $\mathbf{D}(l+1,\cdot)$. Assume that we want to solve $\mathcal{P}(l+1, k)$, and also assume that we fix the last polynomial degree $d_{l+1}$ to $d$. Since the $A_{l+1} E_{l+1}(d_{l+1})$ term is constant, the target function $\sum_{i=1}^{l+1} A_i E_i(d_i)=\sum_{i=1}^{l} A_i E_i(d_i) + A_{l+1}E_{l+1}(d)$ can be replaced to $\sum_{i=1}^{l} A_i E_i(d_i)$. Also, since the $\tau_{l+1}(d_{l+1})$ term is constant, the constraint $\tau(\mathbf{d})=\sum_{i=1}^{l+1}\tau_i(d_i)\le k$ can be regarded as $\sum_{i=1}^{l}\tau_i(d_i)\le k-\tau_{l+1}(d)$. Let us denote $\mathbf{d}'$ in $\mathcal{P}(l+1, k)$ as $(d_1, \cdots, d_l)$, the polynomial degree vector except for the last degree $d_{l+1}$ from $\mathbf{d}=(d_1, \cdots, d_{l+1})$, and the problem $\mathcal{P}(l+1,k)$ with fixed $d_{l+1} = d$ is equivalent to this following problem,
\begin{align}
    \min_{d_1,\cdots, d_{l}\in\mathcal{S}}  \:  V(\mathbf{d}') = \sum_{i=1}^{l} A_i E_i(d_i)\quad \nonumber \text{subject to} \quad \tau(\mathbf{d}')=\sum_{i=1}^{l} \tau_{i}(d_i)\leq k - \tau_{l+1}(d),
\end{align}
where the problem can be regarded as $\mathcal{P}(l,k-\tau_{l+1}(d))$. Since the optimal degree vector for this problem is $\mathbf{D}(l, k-\tau_{l+1}(d))$, the minimum $V(\mathbf{d})$ of the problem $\mathcal{P}(l+1,k)$ with fixed $d_{l+1}=d$ is $V(\mathbf{D}(l,k-\tau_{l+1}(d)) + A_{l+1}E_{l+1}(d) = \sum_{i=1}^l A_i E_i(D_i(l, k-\tau_{l+1}(d))) + A_{l+1}E_{l+1}(d)$. 

Note that there are not so many candidates for $d_{l+1}$ in $\mathcal{P}(l+1,k)$. Since we have a trivial constraint $\tau_{l+1}(d_{l+1})\le k$ in $\mathcal{P}(l+1,k)$, we can investigate all the solutions of $\mathcal{P}(l+1,k)$ with each $d\in \mathcal{S}\cap\{d'\in \mathbb{N}: \tau_{l+1}(d') \le k\}$. Then the solution for $\mathcal{P}(l+1,k)$ is the minimum value of $V(\mathbf{D}(l,k-\tau_{l+1}(d)) + A_{l+1}E_{l+1}(d)$ among these solutions. The value $D_{l+1}(l+1,k)$ is the corresponding polynomial degree $d$ for this minimum value, and $\mathbf{D}(l+1, k) = (\mathbf{D}(l, k-D_{l+1}(l+1,k)),D_{l+1}(l+1,k))$.

Thus, we end up setting the relation between $\mathcal{P}(l+1, k)$ and $\{\mathcal{P}(l, k')\}$ with $k'=1, \cdots, k$. This relation is formalized in the following theorem.

\begin{theorem}
\label{thm:dy}
    For positive $l$ and $k$, let $d'$ be
    \begin{align*}
        \argmin_{d\in\mathcal{S}, \tau_{l+1}(d)\le k} \left[ \sum_{i=1}^l A_i E_i(D_i(l,k- \tau_{l+1} (d))) + A_{l+1}E_{l+1}(d) \right].
    \end{align*}
    Then, $D_i(l+1,k)=D_i(l,k-\tau_{l+1}(d'))$ for each $1\leq i \leq l$, and $D_{l+1}(l+1,k)=d'$.
\end{theorem}

\begin{proof}
To show $D_{i}(l+1,k)=D_i(l,k-\tau_{l+1}(d'))$ for $i=1,\cdots,l$ and $D_{l+1}(l+1,k)=d'$, we have to check that
\begin{align}
\label{eq:dy_res}
\sum_{i=1}^l A_i E_i(D_i(l,k-\tau_{l+1}(d'))) + A_{l+1}E_{l+1}(d') \leq 
\sum_{i=1}^{l+1} A_i E_i (d_i)
\end{align}
for any $d_i\in\mathcal{S}$ which satisfies $\sum_{i=1}^{l+1} \tau_{i}(d_i)\leq k$. Since $\sum_{i=1}^{l} \tau_{i}(d_i)\leq k - \tau_{l+1}(d_{l+1})$, 
\begin{align}
\label{eq:dy_res1}
    \sum_{i=1}^l A_i E_i (D_{i}(l, k-\tau_{l+1}(d_{l+1}))) \leq \sum_{i=1}^l A_i E_i (d_i)
\end{align}
holds for any $d_i\in\mathcal{S}$, from the optimality of $\mathbf{D}(l,k-\tau_{l+1}(d_{l+1}))$. Also, from the definition of $d'$,
\begin{align}
\label{eq:dy_res2}
\begin{aligned}
    &\sum_{i=1}^l A_i E_i(D_i(l,k-\tau_{l+1}(d'))) + A_{l+1}E_{l+1}(d') \\
    &\leq \sum_{i=1}^l A_i E_i(D_i(l,k-\tau_{l+1}(d_{l+1}))) + A_{l+1}E_{l+1}(d_{l+1})
    \end{aligned}
\end{align}
for any $d_{l+1}\in \mathcal{S}$. Combining (\ref{eq:dy_res1}) and (\ref{eq:dy_res2}) leads to (\ref{eq:dy_res}), and this proves the theorem. 
\end{proof}

Theorem \ref{thm:dy} states that the optimal solution $\mathbf{D}(l+1,k)$ of the problem $\mathcal{P}(l+1,k)$ can be efficiently constructed if $\mathbf{D}(l,k')$ has been determined for all $k'=1, \cdots, k$.

\subsection{Dynamic Programming Algorithm}

Based on Theorem \ref{thm:dy}, we construct a dynamic programming algorithm to obtain $\mathbf{D}(N_L,N_K)$. We set an empty two-dimensional table with $N_L$ rows and $N_K$ columns, and each cell is expected to contain $\mathbf{D}(l, k)$ and $V(\mathbf{D}(l, k))$. Each row is sequentially filled in as follows:
\begin{enumerate}
    \item Each cell in the first row in the table is the solution for $\mathcal{P}(1, k)$ for each $k$. Since $E_1$ is a decreasing function in terms of the polynomial degree, the optimal degree is $D_1(1,k)=\max\{d\in\mathcal{S}|\tau_1(d)\leq k\}$ and minimized average loss function is $V(D_1(1, k))=A_1 E_1(D_1(1, k))$ for each $k$.
    \item For fixed $1\leq l<N_L$, assume that $\mathbf{D}(l,k')$'s and $V(\mathbf{D}(l, k'))$ are already known for all $k'=1, \cdots, N_K$. Then, for given $k$, find $d=d(k)$ which minimizes $V(\mathbf{D}(l, k - \tau_{l+1}(d)))  + A_{l+1}E_{l+1}(d)$
    among $d\in\mathcal{S}$ such that $\tau_{l+1}(d)\le k$. Here, the values of $E_i(\cdot)$ have to be pre-computed.
    \item After finding $d(k)$ in the previous step for every $k=1, \cdots, N_K$, we set $\mathbf{D}(l+1,k)$ as $D_i(l+1,k)=D_i(l,k-\tau_{l+1}(d(k)))$ for $1\leq i \leq l$, and $D_{l+1}(l+1,k)=d(k)$. We also set $V(\mathbf{D}(l+1, k)) = V(\mathbf{D}(l, k - \tau_{l+1}(d(k)))) + A_{l+1}E_{l+1}(d)$. This provides the optimal solution $\mathbf{D}(l+1,k)$ and $V(\mathbf{D}(l+1,k))$ for all $k$.
    \item Continue to repeat steps 2 and 3 for $l=1,2,\cdots,N_L-1$ sequentially until we obtain the optimal solution $\mathbf{D}(N_L,N_K)$, which is the degree set in the last cell in the table.
\end{enumerate}

\vspace{-15pt}
\begin{center} 
\begin{minipage}[b]{\linewidth}
\begin{algorithm}[H]
\caption{Dynamic programming algorithm for $\mathcal{P}(L,K)$}\label{alg:cap}
\begin{algorithmic}[1]
\Require The number of layers $N_L$, the scaled integer value of runtime bound $N_K$, the degree space $\mathcal{S}$, the MSE function $E_{i}(\cdot)$, and the discretized runtime function $\tau_i(\cdot)$
\Ensure The optimal degree set $\mathbf{D}(N_L,N_K)$ for the problem $\mathcal{P}(N_L,N_K)$
%\State $l \gets 1$
%\State $k \gets 1/\nu$
\For{$k=1, \cdots, N_K$}
\State $D_1(1,k) \gets \max\{d\in\mathcal{S}|\tau_{1}(d)\leq k\}$
\State $\mathbf{D}(1, k) \gets (D_1(1, k))$
\State $V(1, k)\gets A_1E_1(D_1(1, k))$
\EndFor
\For{$l=1,\cdots,N_L-1$}
\For{$k=1, \cdots, N_K$}
\State 
$d' \gets \argmin_{d\in\mathcal{S}} (V(l, k-\tau_{l+1}(d))  + A_{l+1}E_{l+1}(d))$
\State $\mathbf{D}(l+1,k)\gets(\mathbf{D}(l, k-\tau_{l+1}(d')), d')$
\State $V(l+1, k)\gets V(l, k-\tau_{l+1}(d') + A_{l+1}E_{l+1}(d')$
\EndFor

\EndFor

\State \Return $\mathbf{D}(N_L, N_K)$
\end{algorithmic}
\end{algorithm}
\end{minipage}
\end{center}
\vspace{-5pt}

Note that the proposed algorithm can be executed with a time complexity $O(N_L N_K|\mathcal{S}|)$, which is significantly more efficient than the brute-force searching through all degree search space, which requires a time complexity of $O(|\mathcal{S}|^{N_L})$. The procedure of this algorithm is summarized in Algorithm \ref{alg:cap}.

\section{Optimized Layerwise Approximation Framework}
\label{sec:OLA}

In this section, we specify our optimized layerwise approximation (OLA) framework to apply the generalized distribution-aware approximation $P_{\phi}[d;f](x)$ and the optimized degree from the result of dynamic programming $\mathbf{D}(N_L,N_K)$ in practical inference. 
We outline the entire OLA method in Section \ref{subsec:flow}. 
Additionally, we provide detailed explanations on how the degree search space $\mathcal{S}$ and discretization unit $\nu$ are determined in Section \ref{subsec:searchspace}. We also provide the fine-tuning of the approximate polynomial, which we refer to as the \textit{scaled distribution-aware approximation technique}, in Section \ref{subsec:scaledapp}.

\subsection{The Flow of the OLA Framework}
\label{subsec:flow}
The OLA framework can be executed when the pre-trained model and the dataset used for training are provided. The steps of the framework are as follows:
\begin{enumerate}
    \item \textbf{Determine the input distribution for each layer and average loss variance.} This is the first step to construct an optimization problem. Let $\mu_i$ and $\sigma_i^2$ be set as the mean and the variance of all possible inputs of $i$th activation function for the entire training dataset, respectively. Then, the input distribution for $i$th activation function $\phi_i(x)$ can be approximated as $\mathcal{N}(\mu_i ,\sigma_i^2)$. To determine the average loss variance function $V(\mathbf{d})=\sum_{i=1}^{N_L} A_i E_i(d_i)$, we have to discover the constants $A_i$ and the minimized MSE function $E_i(\cdot)$. The constant $A_i$ can be obtained from the backpropagation process, and $E_i(\cdot)$ follows from Theorem \ref{thm:normal}.

    \item \textbf{Discretize the time latency.} To achieve this, we first obtain the actual time latency $T_i(\cdot)$ for each layer: the runtime of polynomial evaluation $T_1(\cdot)$ and including bootstrapping $T_i(\cdot)$ ($i\geq 2$). Next, determine the degree search space $\mathcal{S}$ and the discretization unit $\nu$ (see Section \ref{subsec:searchspace} for the details).

    \item 
\textbf{Solve the optimization problem with dynamic programming.} For a given constant $N_K$, the optimal solution $\mathbf{D}(N_L,N_K)$ can be done through Algorithm \ref{alg:cap}. Note that if we perform the Algorithm \ref{alg:cap} once, we also can achieve the solution $\mathbf{D}(N_L, k)$ for arbitrary $k\leq N_K$ (see line 9 in the Algorithm \ref{alg:cap}).

\item \textbf{Find the optimal degree set $\mathbf{d}=\mathbf{D}(N_L,k)$.} We have to find the value of $k$ which shows both less accuracy loss and small time latency $\tau(\mathbf{d})$. We use binary search to find the optimized $k$ within the range where the top-1 accuracy obtained through polynomial approximation from the training dataset does not drop below 1\%. To obtain more accurate classification result, we use the scaled distribution-aware approximation here  (see Section \ref{subsec:scaledapp} for the details).
\end{enumerate}

Note that the degrees $\mathbf{d}=\mathbf{D}(N_L,k)$ achieved by the above process is the optimal degrees $\mathbf{d}$ that minimize $\tau(\mathbf{d})$ across the set of degrees, where the top-1 training accuracy differs by less than 1\%. Since the accuracy provided by this degree set does not significantly differ from the original top-1 training accuracy, we can expect it to yield high accuracy on the test dataset as well. %We demonstrate this in the experimental results of the following section.

\begin{figure}
      \begin{center}   \includegraphics[width=0.7\textwidth]{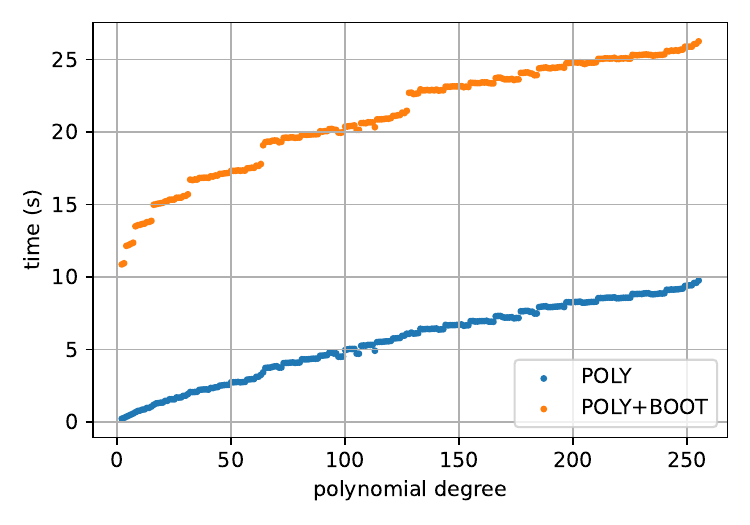}
  \end{center}
  \vspace{-20pt}
  \caption{The runtime of the polynomial evaluation (blue, $T_1(\cdot)$), and the sum of the polynomial evaluation runtime and bootstrapping runtime (orange, $T_i(\cdot)$, $i\geq 2$) on the RNS-CKKS scheme.}
  \label{fig:deg-time}
\end{figure}

\subsection{Setting Degree Search Space and Discretization Unit}
\label{subsec:searchspace}
Considering the time complexity of Algorithm \ref{alg:cap} in solving the optimization problem, which is proportional to the cardinality of $\mathcal{S}$, searching for degrees too finely (e.g., $\mathcal{S}=\{1,2,3,\cdots,1023\}$) might be inefficient. 
We first investigate whether there are degrees with similar time latency $T_i(d)$ in order to remove ``redundant'' degrees from the search space.
Figure \ref{fig:deg-time} shows the exact value of $T_i$'s on the RNS-CKKS scheme with the $\mathtt{Lattigo}$ library \cite{lattigo}.
As shown in the figure, the time latency is almost the same when the polynomial degrees are similar in most cases. 

First, instead of searching every possible degree, we choose $d=2^m-1$ as the candidate degrees to be included in $\mathcal{S}$, for integer $m= 2,3,\cdots,10$.
We note that the maximum polynomial degree that consumes a depth of $m$ on ciphertexts is $2^m-1$, and the bootstrapping performed before polynomial operations varies significantly depending on the depth of the polynomial operations. 
(Check the significant gap between $T_i(2^m-1)$ and $T_i(2^m)$ for $i\geq 2$ in Figure \ref{fig:deg-time}.) 
The accuracy of the approximation polynomial will be almost the same between using a degree of $2^m-1$ and $2^m$, but from a time consumption perspective, using a degree of $2^m-1$ is much more advantageous. 
Therefore, we include $2^m-1$ in $\mathcal{S}$. 

Additionally, as we can see in Figure \ref{fig:deg-time}, there are large gap between $T_i(63)$ and $T_i(127)$, as well as between $T_i(127)$ and $T_i(255)$. Therefore, we include additional degrees in $\mathcal{S}$ for a more fine-grained search, to ensure that the intervals between $T_i(d)$ for 
$d\in\mathcal{S}$ are equally spaced. In our simulation, we set 
\[
\mathcal{S}=\{ 3, 7, 15, 31, 63, 88, 127, 154, 210, 255, 261, 393, 511, 603, 703, 813, 917, 1023\}.
\] 
Also, we choose the discretization unit as $\nu=\frac{1}{4}$ to reduce the loss of values of $T_i$'s due to the discretization. After choosing $\nu$, we can evaluate the discretized time latency,
$\tau(\mathbf{d})=\sum_{i=1}^{N_L}\tau_i(d_i)=\sum_{i=1}^{N_L}\lfloor T_{i}(d_i)/ \nu \rceil$. 

\subsection{Scaled Distribution-Aware Approximation}
\label{subsec:scaledapp}
We identify significant accuracy issues when using distribution-aware approximation in the activation function approximation $P_{\phi}[d;f](x)$ without any modifications. 
%Our OLA method involves approximating using a weighted $l_2$ norm, where the weight distribution is a normal distribution reflecting the average and standard deviation of the input values entering the layer. 
Due to the nature of distribution-aware approximation, the error between the true values is small in regions of high probability but large in regions of low probability. 
Since the probability density function of an input distribution converges rapidly to zero away from the origin in general, the approximation is hardly accurate unless it is close to the origin. 
Therefore, we propose \textit{scaled distribution-aware approximation technique} which finely tunes the approximate polynomials.

There are two main issues with the distribution-aware approximation. One is that the actual distribution of the data might slightly differ from the normal distribution, although it considers the mean and standard deviation of the input data to the layer. In such cases, areas of low probability in the normal distribution may not be so low in the actual data distribution. This discrepancy can significantly affect accuracy if the approximate polynomial fails to accurately approximate the activation function in regions where actual frequent values occur. 
Secondly, even small-probability regions can amplify errors as they pass through the layers, significantly impacting accuracy.
We demonstrate this property through simulations, the details of which are provided in \ref{app:exp}.
Due to these reasons, we experimentally found that using distribution-aware approximation polynomials alone does not allow the neural network to perform meaningful inference.
Thus, there is a need to approximate more accurately, even in low-probability areas, from a conservative perspective.

\begin{figure}
  \begin{center}   \includegraphics[width=0.7\textwidth]{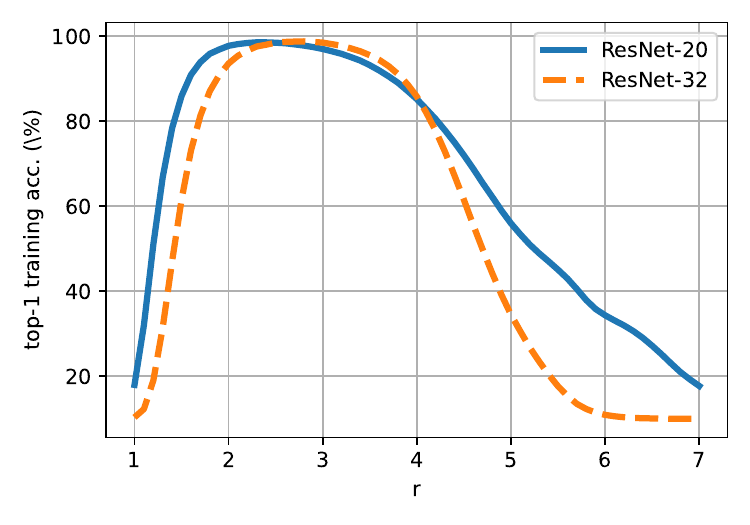}
  \end{center}
  \vspace{-20pt}
  \caption{Accuracy graph as a function of $r$ (with CIFAR-10 dataset).}
  \label{fig:r-vary}
\end{figure}

In the scaled distribution-aware approximation method, we propose to use a new distribution.
We define a distribution $\Phi_i^{[r]}(x)$ for some real value $r \geq 1$ for each layer $i \geq 1$, as $\Phi_i^{[r]}(x):=\frac{1}{r}\phi_i \left( \frac{X-\mu_i}{r}+\mu_i \right)$ where $\mu_i$ is the expectation of the distribution $\phi_i(x)$. 
The transformation $\phi_i \mapsto \Phi_i^{[r]}$ simply \textit{scales} the standard deviation of the original distribution $\phi_i(x)$ by a factor of $r$, while keeping the mean value fixed at $\mu_i$.
Interchanging the weight distribution $\phi_i(x)$ as $\Phi_i^{[r]}(x)$ enables more accurate approximation in low-probability areas of $\Phi_i^{[r]}(x)$ thus mitigating the impacts of distribution differences and error amplification. 
Since these issues about the distribution-aware approximation are related to the whole network, rather than only each layer, we use the same $r$ value for all activation functions in the neural network and control this unified $r$ value. 
Figure \ref{fig:r-vary} shows the effect on the performance of the image classification. 
If we do not apply scaled approximation ($r=1$), we can observe a significantly lower top-1 accuracy obtained from the training dataset. 
By appropriately adjusting the value of $r$, we can find the value that maximizes the training accuracy. 
After finding this $r$ value, we finally adopt the polynomials $\{ P_{\Phi_i^{[r]}}[d_i;f](x)\}_{1\leq i \leq N_L}$ as the approximation polynomial to proceed PI.

\section{Experiment Results for Layerwise Optimization}
\label{sec:result}

In this section, we provide the experimental results, which show the performance of the layerwise optimized approximate polynomials by the proposed algorithm. We evaluate the classification task of the CIFAR-10/100 datasets \cite{krizhevsky2009learning} \bl{and the ImageNet dataset \cite{russakovsky2015imagenet}} with pre-trained ResNet \cite{he2016deep} and ConvNeXt \cite{liu2022convnet} models that incorporate the representative non-arithmetic functions such as ReLU and GELU \cite{hendrycks2016gaussian}, respectively. Also, we analyze the performance of our proposed method and compare it to \bl{the previous state-of-the-art PTA method~\cite{strided}.}

\subsection{Results for the Proposed Approximation Method on the Plaintexts}

\bl{First, to verify the effectiveness of the proposed method, we conduct experiments comparing the performance of previous approximation methods for PTA scenario, the composite minimax approximation \cite{strided} (only available for ReLU) on different models in plaintext.}

\begin{table}
\centering
%\caption{The simulation results on the plaintext CIFAR-10/100 for comparison of $\tau(\mathbf{d})$—an indicator of the inference time on ciphertext and top-1 test accuracy for the obtained optimal degree $\mathbf{d}$ by each approximation method.}
\caption{\bl{The simulation results on the plaintext CIFAR-10/100 and ImageNet datasets for comparison of $\tau(\mathbf{d})$---an indicator of the inference time on ciphertext---and top-1 test accuracy for the obtained optimal degree $\mathbf{d}$ by each approximation method. ``Minimax'' refers to the prior state-of-the art PTA method~\cite{strided}, while ``OLA'' denotes our proposed OLA framework. For ConvNeXt models, Minimax results are not reported since ConvNeXt does not employ ReLU activations.}}

\label{tab:main_result}
\begin{adjustbox}{width=\columnwidth,center}
\begin{tabular}{cccccrl}
\toprule
Backbone                  & Dataset                    & \begin{tabular}[c]{@{}c@{}}Model\\ (Top-1)\end{tabular}                   & \multicolumn{1}{c}{\begin{tabular}[c]{@{}c@{}}PTA\\ Method\end{tabular}}           & \multicolumn{1}{c}{\begin{tabular}[c]{@{}c@{}}Approx.\\ Top-1 (\%)\end{tabular}} & \multicolumn{1}{c}{$\tau(\mathbf{d})$} & \multicolumn{1}{c}{$\mathbf{d}$}                                                                                                    \\ \hline \hline
\multirow{27}{*}{ResNet}  & \multirow{20}{*}{CIFAR-10} & \multirow{2}{*}{\begin{tabular}[c]{@{}c@{}}ResNet-20\\ (91.52\%)\end{tabular}} & Minimax   & 91.55& 2,788 &  All 6,076 \\ \cline{4-7} &&& OLA    & 90.69& 1,106   & 88 31 31 31 31 31 15 31 31 15 31 15 31 31 31 31 31 15 7  \\ 

\cline{3-7}  & & \multirow{3}{*}{\begin{tabular}[c]{@{}c@{}}ResNet-32\\ (92.49\%)\end{tabular}}   & Minimax   & 92.45  & 4,624& All 6,076 \\ \cline{4-7}  & & & OLA   & 91.69  & 1,927  & \begin{tabular}[c]{@{}l@{}}154 63 63 31 63 31 31 15 31 31 31 31 63 15 63 15 63 15 31 15 \\ 31 31 63 31 63 31 31 31 31 31 7\end{tabular}   \\ 

\cline{3-7}   &   & \multirow{4}{*}{\begin{tabular}[c]{@{}c@{}}ResNet-44\\ (92.76\%)\end{tabular}} & Minimax              &  92.66 &6,460  &  All 6,076 \\ \cline{4-7} & &    & OLA    & 91.82    & 3,195                     & \begin{tabular}[c]{@{}l@{}}393 210 154 127 127 63 127 63 88 31 88 31 88 63 88 63 127 31 127 31 \\
127 31 88 31 88 31 88 31 88 63 127 63 127 63 88 63 63 63 63 63 \\
63 31 15\end{tabular}        

\\ \cline{3-7} &&  \multirow{4}{*}{\begin{tabular}[c]{@{}c@{}}ResNet-56\\ (93.27\%)\end{tabular}}    & Minimax              & 93.12  & 8,296 & All 6,076   \\ \cline{4-7} &&& OLA  & 92.29 & 3,879 & \begin{tabular}[c]{@{}l@{}}255 63 210 88 127 63 127 31 88 31 88 31 63 63 63 31 63 31 63 63 \\
127 31 127 31 127 31 88 31 63 31 63 31 63 31 63 31 63 63 127 63 \\
88 63 88 63 63 63 63 31 63 31 63 31 31 31 15\end{tabular}             \\

\cline{3-7} && \multirow{7}{*}{\begin{tabular}[c]{@{}c@{}}ResNet-110\\ (93.50\%)\end{tabular}} & Minimax              & 93.44 &16,558 &All 6,076 \\ \cline{4-7}&&&OLA                  & 92.56                                                                            & 9,448                     & \begin{tabular}[c]{@{}l@{}}917 255 511 127 511 255 255 127 255 88 255 63 210 63 210 63 255 63 210 63 \\
210 63 210 63 210 63 210 63 210 63 210 63 154 63 210 63 210 88 255 31 \\
255 31 255 31 255 31 255 63 255 63 255 63 255 63 255 63 255 63 255 31 \\
255 63 255 63 255 63 210 63 210 63 210 63 255 127 255 63 255 63 255 63 \\
255 63 210 63 210 63 210 63 210 63 210 63 210 63 127  63 127 63 127 63 \\
127 63 127 63 127 63 63 63 63\end{tabular} \\

\cline{2-7} & \multirow{5}{*}{CIFAR-100} & \multirow{2}{*}{\begin{tabular}[c]{@{}c@{}}ResNet-20\\ (63.08\%)\end{tabular}}                                        & Minimax            &63.06&2,788&All 6,076\\ \cline{4-7}&&&OLA                  & 62.76                                                                             &1,593                   & 393 127 127 63 127 88 127 88 127 63 63 63 63 127 210 127 210 127 51   \\ \cline{3-7} 
&                            & \multirow{3}{*}{\begin{tabular}[c]{@{}c@{}}ResNet-32\\ (64.61\%)\end{tabular}}& Minimax &63.92&4,624&All 6,076\\ \cline{4-7} 
&                            &                                                                                & OLA                  & 64.06                                                                            & 2,951                     & \begin{tabular}[c]{@{}l@{}}511 255 210 154 210 127 210 88 210 88 210 210 210 127 210 127 210 88 210 127  \\ 210 210 255 210 255 210 255 210 255 210 210 \end{tabular}                                       \\ \cline{2-7} 
& \multirow{2}{*}{ImageNet}  & \multirow{2}{*}{{\begin{tabular}[c]{@{}c@{}}ResNet-18\\ (71.47\%)\end{tabular}}}                                                     & Minimax           & 71.41 & 2,482& All 6,076\\ \cline{4-7} 
&&& OLA                  & 70.58                                                                            & 1,753                     & 703 511 511 255 255 255 255 210 255 210 255 210 210 210 154 255 63                                       
\\ \hline
\multirow{14}{*}{ConvNeXt} & \multirow{5}{*}{CIFAR-10}  & \begin{tabular}[c]{@{}c@{}}ConvNeXt-T\\ (92.16\%)\end{tabular}                 & \multirow{14}{*}{OLA} & 91.42                                                                            & 749                      & 3 3 3 3 3 3 3 3 3 3 3 3 3 3 3 3 3 3                                                                                                                                                                                                                                                                                                                                                                                                                 \\ \cline{3-3} \cline{5-7} 
&                            & \begin{tabular}[c]{@{}c@{}}ConvNeXt-S\\ (92.85\%)\end{tabular} &                      & 91.93   & 1,541                     & 3 3 3 3 3 3 3 3 3 3 3 3 3 3 3 3 3 3 3 3 3 3 3 3 3 3 3 3 3 3 3 3 3 3 3 3                                          
\\ \cline{3-3} \cline{5-7} &&\begin{tabular}[c]{@{}c@{}}ConvNeXt-B\\ (93.01\%)\end{tabular}  && 92.32                                                                            & 1,569                     & 31 3 7 3 7 3 3 3 3 3 3 3 3 3 3 3 3 3 3 3 3 3 3 3 3 3 3 3 7 7 3 3 3 3 3 3                                                             \\ \cline{2-3} \cline{5-7} & \multirow{3}{*}{CIFAR-100} & \begin{tabular}[c]{@{}c@{}}ConvNeXt-T\\ (70.20\%)\end{tabular}                                                                   &                      & 69.78                                                                            & 788                      & 15 7 7 3 7 3 3 3 3 3 3 7 3 3 7 3 7 7                                                                                                                      \\ \cline{3-3} \cline{5-7}&                            & \begin{tabular}[c]{@{}c@{}}ConvNeXt-S\\ (70.38\%)\end{tabular}&                      & 69.75                                                                            & 1,595                     & 31 7 7 3 3 3 3 7 3 3 3 3 7 3 3 7 3 3 3 3 3 3 3 7 3 3 3 3 3 3 3 7 3 3 3 15                                                                                        \\ \cline{2-3} \cline{5-7} 
& \multirow{3}{*}{ImageNet}  & \begin{tabular}[c]{@{}c@{}}ConvNeXt-T\\ (83.96\%)\end{tabular}                                                        &                      & 83.08                                                                           & 943                      & 31 15 15 15 15 15 15 15 15 15 15 15 15 15 15 15 15 15  \\ \cline{3-3} \cline{5-7} 
&                            & \begin{tabular}[c]{@{}c@{}}ConvNeXt-S\\ (84.89\%)\end{tabular}                                                                     &                      & 84.64                                                                           & 1,597                     & \begin{tabular}[c]{@{}l@{}}31 15 15 15 15 15 15 31 15 15 15 15 15 15 15 15 31 31 15 15 15\\ 15 15 15 15 15 15 15 15 15 15 15 15 15 15 15\end{tabular}                                     \\ 
\bottomrule
\end{tabular}
\end{adjustbox}
\end{table}

\bl{The results are shown in Table \ref{tab:main_result}. We observe that the OLA framework consistently achieves lower $\tau(\mathbf{d})$ than the composite minimax approximation across all ResNet architectures. Furthermore, our simulations demonstrate that for ConvNeXt models with the GELU activation function, effective inference can be carried out with low-degree polynomial approximations without retraining or modifying the original models. On CIFAR-10, satisfactory classification performance was obtained with only a cubic polynomial, while on large-scale datasets such as ImageNet, inference was successfully performed with polynomials of degree no greater than 31, achieving a top-1 accuracy of 84.64\%.}

\bl{Furthermore, we conducted an ablation study to examine the contribution of each component of the OLA framework to the overall performance, and the results are summarized in Table \ref{tab:ablation}. The core of OLA lies in (i) distribution-aware approximation based on layerwise input distributions and (ii) dynamic programming for adjusting polynomial degrees per layer. Using the weighted least square method to approximate non-arithmetic activations, we tested the presence or absence of these two factors and report the resulting performance. Table \ref{tab:ablation} compares different approximation methods on ResNet models trained on the CIFAR-10 dataset. As can be seen from the table, the full OLA framework, which applies both distribution-aware approximation and dynamic programming, achieved the lowest $\tau(\mathbf{d})$. In particular, distribution-aware approximation alone already reduced the latency significantly, and adding dynamic programming further decreased $\tau(\mathbf{d})$. In ResNet-20, the reduction from dynamic programming was modest (1,142 $\rightarrow$ 1,106, a 3.15\% decrease), whereas in ResNet-110, the reduction was substantial (10,725 $\rightarrow$ 9,448, a 11.91\% decrease), confirming that the effect of dynamic programming becomes more pronounced as the network depth increases.
}

\begin{table}
\centering
\caption{\bl{Ablation study on the proposed OLA framework, analyzing the impact of distribution-aware approximation and dynamic programming on ResNet models with the plaintext CIFAR-10 dataset. The mark \XSolidBrush\ indicates that distribution-aware approximation is replaced by weighted least square without considering input distributions, and that dynamic programming (DP) is disabled by assigning the same degree to all layers.}
}
\label{tab:ablation}
\begin{adjustbox}{width=\columnwidth,center}

\begin{tabular}{ccccrl}
\toprule[1.5pt]
Backbone                                                                        & Dist.                       & DP                          & Top-1 (\%) & \multicolumn{1}{c}{$\tau(\mathbf{d})$} & \multicolumn{1}{c}{$\mathbf{d}$}                                                                                                                                                                                                                                                                                                                                                                                                                                                                    \\              \hline \hline
\multirow{4}{*}{\begin{tabular}[c]{@{}c@{}}ResNet-20\\ (91.52\%)\end{tabular}}  & \XSolidBrush & \XSolidBrush & 91.06      & 1,440                                  & All 88                                                                                                                                                                                                                                                                                                                                                                                                                                                                                                           \\ \cline{2-6} 
                                                                                & \XSolidBrush & \Checkmark   & 90.79      & 1,238                                  & 154 63 63 63 63 63 63 63 63 31 63 31 31 63 31 63 31 63 3                                                                                                                                                                                                                                                                                                                                                                                                                                                         \\ \cline{2-6} 
                                                                                & \Checkmark   & \XSolidBrush & 90.99      & 1,142                                  & All 31                                                                                                                                                                                                                                                                                                                                                                                                                                                                                                           \\ \cline{2-6} 
                                                                                & \Checkmark   & \Checkmark   & 90.69      & \textbf{1,106}        & 88 31 31 31 31 31 15 31 31 15 31 15 31 31 31 31 31 15 7                                                                                                                                                                                                                                                                                                                                                                                                                                                          \\ \toprule[1.5pt]
\multirow{4}{*}{\begin{tabular}[c]{@{}c@{}}ResNet-32\\ (92.49\%)\end{tabular}}  & \XSolidBrush & \XSolidBrush & 91.73      & 2,388                                  & All 88                                                                                                                                                                                                                                                                                                                                                                                                                                                                                                           \\ \cline{2-6} 
                                                                                & \XSolidBrush & \Checkmark   & 91.50      & 2,286                                  & 255 127 127 88 88 63 88 63 88 63 88 63 88 63 88 63 63 63 63 63 63 127 63 127 63 127 63 88 63 63 7                                                                                                                                                                                                                                                                                                                                                                                                                \\ \cline{2-6} 
                                                                                & \Checkmark   & \XSolidBrush & 91.53      & 2,143                                  & All 63                                                                                                                                                                                                                                                                                                                                                                                                                                                                                                           \\ \cline{2-6} 
                                                                                & \Checkmark   & \Checkmark   & 91.69      & \textbf{1,927}        & 154 63 63 31 63 31 31 15 31 31 31 31 63 15 63 15 63 15 31 15 31 31 63 31 63 31 31 31 31 31 7                                                                                                                                                                                                                                                                                                                                                                                                                     \\ \toprule[1.5pt]
\multirow{4}{*}{\begin{tabular}[c]{@{}c@{}}ResNet-44\\ (92.76\%)\end{tabular}}  & \XSolidBrush & \XSolidBrush & 92.05      & 4,191                                  & All 210                                                                                                                                                                                                                                                                                                                                                                                                                                                                                                          \\ \cline{2-6} 
                                                                                & \XSolidBrush & \Checkmark   & 91.81      & 4,145                                  & \begin{tabular}[c]{@{}l@{}}703 511 511 255 255 127 255 210 210 127 210 127 210 210 210 210 210 154 210 154\\ 210 154 210 127 210 127 210 127 210 255 210 255 210 255 154 255 127 255 127 255 63 127 15\end{tabular}                                                                                                                                                                                                                                                                                              \\ \cline{2-6} 
                                                                                & \Checkmark   & \XSolidBrush & 91.96      & 3,336                                  & All 88                                                                                                                                                                                                                                                                                                                                                                                                                                                                                                           \\ \cline{2-6} 
                                                                                & \Checkmark   & \Checkmark   & 91.82      & \textbf{3,195}        & \begin{tabular}[c]{@{}l@{}}393 210 154 127 127 63 127 63 88 31 88 31 88 63 88 63 127 31 127 31 \\ 127 31 88 31 88 31 88 31 88 63 127 63 127 63 88 63 63 63 63 63 63 31 15\end{tabular}                                                                                                                                                                                                                                                                                                                           \\ \toprule[1.5pt]
\multirow{4}{*}{\begin{tabular}[c]{@{}c@{}}ResNet-56\\ (93.27\%)\end{tabular}}  & \XSolidBrush & \XSolidBrush & 92.57      & 7,026                                  & All 393                                                                                                                                                                                                                                                                                                                                                                                                                                                                                                          \\ \cline{2-6} 
                                                                                & \XSolidBrush & \Checkmark   & 92.44      & 6,222                                  & \begin{tabular}[c]{@{}l@{}}917 511 511 511 511 511 511 255 511 255 511 255 511 255 255 255 255 255 255 511\\ 511 255 511 255 255 255 255 255 255 255 255 255 255 210 255 255 255 511 255 511\\ 255 511 255 511 255 511 210 511 210 511 127 255 88 210 31\end{tabular}                                                                                                                                                                                                                                            \\ \cline{2-6} 
                                                                                & \Checkmark   & \XSolidBrush & 92.62      & 4,284                                  & All 88                                                                                                                                                                                                                                                                                                                                                                                                                                                                                                           \\ \cline{2-6} 
                                                                                & \Checkmark   & \Checkmark   & 92.29      & \textbf{3,879}         & \begin{tabular}[c]{@{}l@{}}255 63 210 88 127 63 127 31 88 31 88 31 63 63 63 31 63 31 63 63 \\ 127 31 127 31 127 31 88 31 63 31 63 31 63 31 63 31 63 63 127 63 \\ 88 63 88 63 63 63 63 31 63 31 63 31 31 31 15\end{tabular}                                                                                                                                                                                                                                                                                       \\ \toprule[1.5pt]
\multirow{4}{*}{\begin{tabular}[c]{@{}c@{}}ResNet-110\\ (93.50\%)\end{tabular}} & \XSolidBrush & \XSolidBrush & 92.51      & 21,172                                 & All 917                                                                                                                                                                                                                                                                                                                                                                                                                                                                                                          \\ \cline{2-6} 
                                                                                & \XSolidBrush & \Checkmark   & 92.56      & 17,339                                 & \begin{tabular}[c]{@{}l@{}}1023 1023 1023 1023 1023 1023 1023 1023 1023 917 917 813 813 703 813 813 813 511 813 511 \\ 813 511 813 511 703 511 703 511 703 511 511 511 511 511 703 703 813 917 813 511 \\ 813 511 813 511 813 511 813 511 813 511 813 511 703 511 703 511 511 511 511 511 \\ 511 511 511 511 511 511 511 511 511 511 511 813 511 1023 511 917 511 917 511 917 \\ 511 917 511 917 511 917 511 917 511 917 511 917 511 813 511 813 511 813 511 813 \\ 511 703 511 813 511 511 255 511\end{tabular} \\ \cline{2-6} 
                                                                                & \Checkmark   & \XSolidBrush & 92.66      & 10,725                                 & All 210                                                                                                                                                                                                                                                                                                                                                                                                                                                                                                          \\ \cline{2-6} 
                                                                                & \Checkmark   & \Checkmark   & 92.56      & \textbf{9,448}        & \begin{tabular}[c]{@{}l@{}}917 255 511 127 511 255 255 127 255 88 255 63 210 63 210 63 255 63 210 63 \\ 210 63 210 63 210 63 210 63 210 63 210 63 154 63 210 63 210 88 255 31 \\ 255 31 255 31 255 31 255 63 255 63 255 63 255 63 255 63 255 63 255 31 \\ 255 63 255 63 255 63 210 63 210 63 210 63 255 127 255 63   255 63 255 63 \\ 255 63 210 63 210 63 210 63 210 63 210 63 210 63 127 63 127 63 127 63 \\ 127 63 127 63 127 63 63 63 63\end{tabular}                                                   \\
                                                                                \bottomrule[1.5pt]
\end{tabular}

\end{adjustbox}
\end{table}

\subsection{Results for the Proposed Method on the Ciphertexts}

We use the implementation in \cite{strided} as our baseline to validate the proposed method empirically. We set the degree of polynomial for ciphertexts $N=16$. As the previous work used the total modulus bit length $\log(PQ)=1553$ with the secret key of Hamming weight $h=192$ for satisfying the 128-bit security, we also optimized the modulus bit length of the proposed moduli-chain. In the same way as the previous work, we set base modulus, bootstrapping moduli, and special moduli as 51-bit primes and other evaluation moduli as 46-bit primes. For each implementation, the number of special moduli is determined to satisfy 128-bit security while achieving the fastest key switching time. For other parameters, such as bootstrapping parameters, we follow the settings used in \cite{strided}.

%Our approximation techniques are independent of the convolution operation for FHE and can be applied to alternative implementations (e.g., \cite{kim2023optimized, cheon2024batch}) for similar benefits. 

\begin{table}
\caption{The top-1 accuracy and the time latency of ResNet models on the homomorphically encrypted CIFAR-10 by the Lattigo library.}

\label{tab:acc}

\begin{center}
\begin{adjustbox}{width=0.75\columnwidth,center}
\begin{tabular}{ccccc}
\toprule
\multirow{2}{*}{Backbone}  &\multicolumn{2}{c}{Minimax \cite{strided}}&\multicolumn{2}{c}{OLA (Ours)}
\\ 
&Accuracy & Time (s) &Accuracy & Time (s) 
\\\hline 
ResNet-20 (91.92\%)&91.31\% &1,231&90.55\%&\textbf{407}($\times$3.02$\downarrow$)\\
ResNet-32 (92.83\%)&92.40\% &1,913&91.67\%&\textbf{679}($\times$2.82$\downarrow$)\\ 
\bottomrule
\end{tabular}
\end{adjustbox}
\end{center}

\end{table}

Under the same condition of 128-bit security, we compare the previous state-of-the-art approximation method on PTA and the proposed method to show how much inference runtime is saved by the proposed method. Table \ref{tab:acc} shows the results. The proposed optimization leads to a significant reduction in total inference time for the ResNet-32 models compared to using the minimax approximation.
We confirm that the proposed method can construct accurate models with reduced polynomial degrees while incurring a marginal reduction in classification accuracy for ciphertext as well.

\subsection{Comparison of the Performed Simulation with Previous Works}

Similar to PTA work that leverages pre-trained models, research using AAT for PI has also been actively developed recently. In addition to the polynomial approximation method mentioned in Table \ref{tab:main_result}, there are various available optimization techniques. 
Since each technique is used in different environments and aims at different goals, it is problematic to make a precise quantitative comparison between them. 
We compared the characteristics of recent studies, particularly those based on ResNets, with a similar number of layers (ResNet-18 or 20), in Table \ref{tab:comparison}.

\begin{table}
\caption{Comparison of recent PI works and our work using ResNet-18/20 as the backbone. The asterisk (*) indicates that the accuracy is achieved on a partial test dataset. The double asterisk (**) indicates that the value is estimated using the information from DELPHI \cite{mishra2020delphi}.}
\label{tab:comparison}
\begin{center}
\begin{adjustbox}{width=\columnwidth,center}

\begin{tabular}{cccccccc}
\toprule
Area&Dataset&Study&Scheme&Retraining&Latency(s)& Comm. (MB)& Acc. (\%) \\ \hline \hline
AAT & CIFAR-10 & HyPHEN \cite{kim2023hyphen} & RNS-CKKS \cite{RNS-CKKS}  & \Checkmark & 37.57 (64 threads)  & 0&92.17 \\
AAT & CIFAR-100 & DeepReDuce \cite{jha2021deepreduce} & DELPHI \cite{mishra2020delphi} &   \Checkmark &0.455 &25.14${}^{**}$& 65.00 \\
AAT & Tiny-ImageNet & DeepReDuce \cite{jha2021deepreduce} & DELPHI \cite{mishra2020delphi} &  \Checkmark &4.6 &469.8${}^{**}$& 59.18 \\
AAT & ImageNet & HyPHEN \cite{kim2023hyphen} & RNS-CKKS \cite{RNS-CKKS} & \Checkmark & 373.41 (64 threads)&0& 65.25 \\ \hline
PTA & CIFAR-10 &  \cite{lee2022privacy} & RNS-CKKS \cite{RNS-CKKS} & \XSolidBrush & 10,602 & 0 & 92.43${}^{*}$ \\
PTA & CIFAR-10 &  \cite{strided} & RNS-CKKS \cite{RNS-CKKS} & \XSolidBrush & 1,231 & 0 & 91.31 \\
%PTA & CIFAR-10 &  \cite{kim2023optimized} & CKKS \cite{cheon2017homomorphic} & \XSolidBrush & 368 & 0 & 90.32 \\
%PTA & CIFAR-10 &  \cite{cheon2024batch} & RNS-CKKS \cite{RNS-CKKS}  & \XSolidBrush & 8067.90${}^*$ (40 threads)&0& 93.65 \\
PTA & CIFAR-10 &  OLA (Ours) & RNS-CKKS \cite{RNS-CKKS}& \XSolidBrush & 407 &0& 91.67 \\

\bottomrule
\end{tabular}
\end{adjustbox}
\end{center}

\end{table}

In the case of HyPHEN \cite{kim2023hyphen}, significant reductions in inference latency on FHE-encrypted data were achieved by optimizing the convolution method and data packing technique, while approximating the ReLU function with a low-degree polynomial. However, the model had to undergo a retraining process to obtain one that suited the use of the low-degree polynomial.

DeepReDuce \cite{jha2021deepreduce} demonstrates significantly lower time latency compared to other studies. 
This is largely due to its use of multi-party computation (MPC), which distinguishes it from approaches like ours or \cite{kim2023hyphen}. 
Unlike FHE, where the server is responsible for all computations, MPC allows for a wider range of operations to be performed more quickly by enabling secret sharing between the server and the client.
For example, in the MPC protocol DELPHI \cite{mishra2020delphi}, a single ReLU operation can be exactly executed in just 85.3$\mu s$. 
However, while MPC reduces computation time, it incurs a trade-off in the form of communication overhead between the server and client. 
Since \cite{jha2021deepreduce} does not report the exact communication costs, we estimated the overall communication overhead by using the cost per ReLU from DELPHI and the total number of ReLUs in \cite{jha2021deepreduce}.

On the PTA side, prior works \cite{lee2022privacy, lee2023precise, strided} adopted minimax polynomial approximation for ReLU. In particular, \cite{strided} achieved an inference latency of 1,231 seconds (ResNet-20 for CIFAR-10). Our OLA method successfully reduces the latency by nearly threefold.
Additionally, previous studies beyond polynomial approximation in other PTA works, such as optimized convolution methods \cite{kim2023optimized} or batch inference on multiple data points \cite{cheon2024batch}, could be integrated with our OLA method to further reduce inference latency.

Both AAT and PTA are meaningful in the context of PI work, depending on the specific environments in which the PI is performed. 
Our PTA work presents a method that effectively approximates the activation function using polynomials without incurring any communication costs, which is particularly useful when retraining the model may not be feasible.

\subsection{Computational Overhead of OLA Framework}

\bl{Finally, we summarize the time overhead required to determine the approximation polynomials to execute the OLA framework. Table \ref{tab:preprocessing} reports the time needed for profiling the input distribution of each layer for different datasets and backbones, as well as the time for solving the resulting optimization problem using dynamic programming. As can be seen in the table, for CIFAR-10/100 these pre-processing steps were completed within only a few minutes, indicating that the overhead is minor. For ImageNet, the input distribution profiling required about three hours. The difference compared to CIFAR-10/100 is primarily due to the much larger size of the ImageNet training dataset (approximately 138 GB). Importantly, this profiling step needs to be performed only once during polynomial approximation. Once completed, inference can be carried out repeatedly on test data without any additional cost. Therefore, we consider this overhead acceptable in practice.}

\begin{table}
\centering
%\caption{The simulation results on the plaintext CIFAR-10/100 for comparison of $\tau(\mathbf{d})$—an indicator of the inference time on ciphertext and top-1 test accuracy for the obtained optimal degree $\mathbf{d}$ by each approximation method.}
\caption{\bl{The results on the pre-processing overhead of OLA framework, measuring the time cost of input distribution profiling and dynamic programming across various backbones and datasets. The dagger ($\dagger$) indicates dynamic programming result obtained with a smaller degree search space, $\mathcal{S}=\{3,7,15,31,63\}$, than described in Section~\ref{subsec:searchspace}.
}}

\label{tab:preprocessing}
\begin{adjustbox}{width=\columnwidth,center}
\begin{tabular}{ccrrrrrrr}
\toprule
Dataset         & \multicolumn{8}{c}{CIFAR-10}                                                                                                                                                                                                                                    \\ \hline
Backbone        & ResNet-20                 & \multicolumn{1}{c}{ResNet-32} & \multicolumn{1}{c}{ResNet-44}  & \multicolumn{1}{c}{ResNet-56}  & \multicolumn{1}{c}{ResNet-110} & \multicolumn{1}{c}{ConvNeXt-T} & \multicolumn{1}{c}{ConvNeXt-S} & \multicolumn{1}{c}{ConvNeXt-B} \\ \hline
Input Dist. (s) & \multicolumn{1}{r}{8.12}  & 9.29                          & 10.69                          & 11.68                          & 17.97                          & 22.99                          & 38.21                          & 56.18                          \\
Dynamic P. (s)  & \multicolumn{1}{r}{7.76} & 30.91                         & 80.47                          & 159.90                         & 773.07                         & 0.08${}^\dagger$                         & 0.34${}^\dagger$                         & 0.37${}^\dagger$                         \\ \hline \hline
Dataset         & \multicolumn{4}{c}{CIFAR-100}                                                                                               & \multicolumn{1}{c}{}           & \multicolumn{3}{c}{ImageNet}                                                                     \\ \cline{1-5} \cline{7-9} 
Backbone        & ResNet-20                 & \multicolumn{1}{c}{ResNet-32} & \multicolumn{1}{c}{ConvNeXt-T} & \multicolumn{1}{c}{ConvNeXt-S} & \multicolumn{1}{c}{}           & \multicolumn{1}{c}{ResNet-18}  & \multicolumn{1}{c}{ConvNeXt-T} & \multicolumn{1}{c}{ConvNeXt-S} \\ \cline{1-5} \cline{7-9} 
Input Dist. (s) & \multicolumn{1}{r}{8.50}  & 10.01                         & 23.16                          & 38.23                          &                                & 9510.94                        & 9572.22                        &     8973.55                           \\
Dynamic P. (s)  & \multicolumn{1}{r}{8.00}  & 34.30                         & 0.09${}^\dagger$                         & 0.40${}^\dagger$                         &                                & 6.23                           & 0.08${}^\dagger$                         &   0.33${}^\dagger$           \\
\bottomrule
\end{tabular}

\end{adjustbox}
\end{table}

\section{Conclusion}
We discussed the necessity of PPML, focusing on HE for secure cloud-based machine learning by encrypting user data to protect privacy. In the context of PI, the key challenge is approximating non-arithmetic functions as HE supports only basic arithmetic operations. 
This study targeted the PTA approach, aiming to improve the efficiency of approximating activation functions with polynomials compared to the state-of-the-art minimax approximation method. 
Our newly proposed OLA method addresses PTA inefficiencies by customizing polynomial approximations for each neural network layer, significantly improving inference time while maintaining classification accuracy. 
We demonstrated that our proposed approximation method can be successfully applied to standard convolutional neural networks such as ResNet and ConvNeXt. It effectively classifies \bl{CIFAR-10/100 and ImageNet} datasets while significantly reducing time latency compared to previous approaches.

There are still several ways to further reduce time latency when performing PI in the PTA scenario. As the need for securely running large models like large language models (LLMs) continues to grow, the development of PTA methods, which avoid model retraining, becomes even more critical. In Transformer architectures \cite{vaswani2017attention}, which are widely used in LLMs, the softmax function—one of the non-arithmetic functions—is frequently used. Since softmax is a multivariate function, new polynomial approximation techniques need to be developed to handle it efficiently. As future work, we propose exploring polynomial approximations for non-arithmetic multivariate functions and developing new approximation methods tailored for LLMs.

\section*{Acknoledgements}
This work was supported by Institute of Information \& communications Technology Planning \& Evaluation (IITP) grant funded by the Korea government(MSIT) (RS-2024-00399401, Development of Quantum-Safe Infrastructure Migration and Quantum Security Verification Technologies).

\appendix

\section{Preliminaries: RNS-CKKS Scheme and Moduli-Chain Managing}
\label{app:prelim}
In this section, we provide details of the RNS-CKKS scheme used in our simulation, along with the bootstrapping process described in Section \ref{sec:rns}. Additionally, we explain the moduli-chain management method we applied to optimize the bootstrapping latency for private inference in convolutional neural networks in Section \ref{sec:moduli}.
\subsection{RNS-CKKS Scheme}
\label{sec:rns}
The FHE scheme is an encryption scheme that supports operations on encrypted data, and the residue number system variant of Cheon-Kim-Kim-Song (RNS-CKKS) scheme \cite{RNS-CKKS} is one of the prominent FHE schemes that support operations for real (or complex) number data. The RNS-CKKS scheme encrypts real (or complex) number data stored in the form of a 1-dimensional vector into the form $(b,a)\in \mathcal{R}_{Q_l}^2$, where $Q_l:=\prod_{i=0}^l q_i$ is the product of prime numbers $q_i$, and $\mathcal{R}_{Q_l}:=\mathbb{Z}_{Q_l}[X]/\langle X^N + 1 \rangle$. The primes $q_i$ are used to represent the coefficients of elements in $\mathcal{R}_{Q_l}^2$ as a residual number system to accelerate operations among ciphertexts. When a ciphertext is in $\mathcal{R}_{Q_l}^2$, its level is referred to as $l$ ($l\geq 0$), where the ciphertext level refers to computational resources that define the possible number of homomorphic multiplications.

There are five primary homomorphic operations in the RNS-CKKS scheme: addition, multiplication with plaintext, multiplication with ciphertext, rotation, and conjugation. Among these, the homomorphic multiplication with a plaintext or a ciphertext decreases the ciphertext level by one. When the ciphertext reaches level 0, it is no longer possible to perform homomorphic multiplication. Therefore, the server receiving the ciphertext is restricted in using homomorphic multiplication without decryption or the assistance of a third party. However, through bootstrapping, the level of the ciphertext can be increased, allowing further homomorphic multiplications to be performed. 

When we evaluate the polynomial for ciphertexts, we perform a baby-step giant-step (BSGS) algorithm to evaluate the polynomial with optimal depth $\lceil \log_2 (d+1)\rceil$. \cite{bossuat2021efficient} The BSGS algorithm decomposes the polynomial as a quotient polynomial and a remainder polynomial dividing the Chebyshev polynomials. We proactively store the coefficients of those decomposed polynomials in the tree structure. The BSGS algorithm evaluates Chebyshev polynomials from the input values as a baby-step, and combines them using the pre-stored coefficients as a giant-step. 
The one thing we have to consider for performing the BSGS algorithm is that the magnitudes of the coefficients the algorithm stores in the tree should not be excessively large.

Bootstrapping is the process of transforming a ciphertext $\mathsf{ct}$ at level 0, denoted as $\mathsf{ct} \in \mathcal{R}_{Q_0}^2$, into a ciphertext $\mathsf{ct}_{\text{Boot}}$ at level $l' > 0$, represented as $\mathsf{ct}_{\text{Boot}} \in \mathcal{R}_{Q_{l'}}^2$, while ensuring that the decryption results of $\mathsf{ct}$ and $\mathsf{ct}_{\text{Boot}}$ are the same with some approximation noise. To achieve this, we first raise the level of $\mathsf{ct}$ to $L_{\text{Boot}}$, irrespective of the decryption result. The resulting ciphertext $\mathsf{ct}'$ will have a decryption result in the form of $\Delta \cdot m + q_0 \cdot I$ for some integer coefficient polynomial $I$, where $m$ is the message polynomial, decryption of $\mathsf{ct}$. Subsequently, we perform modular reduction to eliminate $q_0 \cdot I$. During this process, the level of the ciphertext is reduced by $l_{\text{Boot}}$, resulting in a ciphertext with a level of $l'=L_{\text{Boot}} - l_{\text{Boot}}$. As a result, the decryption result of the final ciphertext closely matches the decryption result of $\mathsf{ct}$ while having a level of $l'$. To perform multiple operations among ciphertexts using FHE, it is essential to carefully design the entire operation process, taking into consideration factors such as bootstrapping and the consumption of levels.
According to the results of deep neural network operations in the previously conducted research using the RNS-CKKS scheme, bootstrapping consumes more than 70\% of the overall time \cite{strided}. Therefore, while it is possible to use bootstrapping as much as needed to continue homomorphic multiplication, it comes at the cost of significant time consumption. 

\subsection{Moduli-Chain Managing} \label{sec:moduli}
When approximating non-arithmetic operations with polynomials for private inference in deep neural networks on the RNS-CKKS scheme, previous works usually used the same polynomials for each layer. Thus, it is sufficient to use a single moduli-chain for implementing the deep neural network on FHE. However, the proposed polynomial approximation approach involves applying different polynomials with different degrees for each non-arithmetic operation by considering the characteristics of each layer. 
The single moduli-chain in the state-of-the-art scheme is designed to accommodate the maximum depth of the approximate polynomials. 
However, there is a problem of unused levels occurring in the single moduli-chain, when it is applied to the proposed approach. This makes the input data size to bootstrapping larger than optimal and thus noticeably increases the runtime. Therefore, we propose a method to optimize the bootstrapping runtime by the different moduli-chain for each layer as the different depth is used for each activation function.
In this section, we propose a method to reduce the runtime of the bootstrapping by removing unused moduli based on the specific requirements of each layer, thus optimizing the situation for each layer accordingly. 
\begin{figure}[!t]
  \begin{center}   \includegraphics[width=0.6\textwidth]{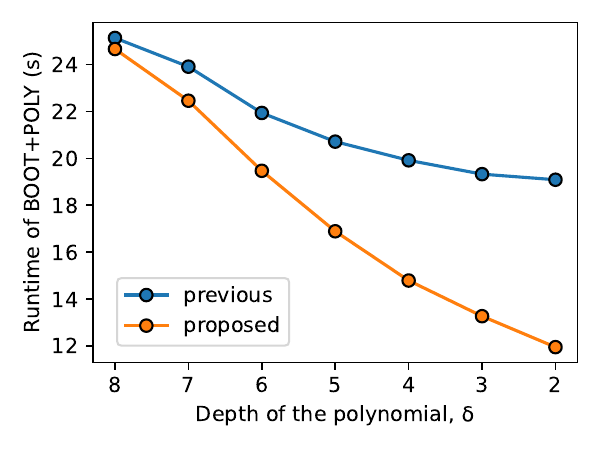}
  \end{center}
  \vspace{-20pt}
  \caption{Comparison of runtime between maintaining the modulus and moduli-chain managing methods, the increasing function of the depth consumption $\delta$.}
  \label{fig:moduli}
\end{figure}
Previous works designed the moduli-chain for private inference on deep neural networks as follows. The modulus of the ciphertext is denoted as $Q_L = \prod_{i=0}^L q_i$. Here, $q_0$ represents the base modulus, and $q_1, \cdots, q_\delta$ correspond to the moduli required for deep neural network operations that consume depth, such as convolutional layers or approximate activation polynomials. The remaining $q_{\delta+1}, \cdots, q_{L}$ are moduli for bootstrapping, each of which is far larger than $q_1, \cdots, q_\delta$. When performing private inference, we use moduli $q_1, \cdots, q_\delta$ to evaluate homomorphic operations. Then, when the level of the ciphertext becomes 0, we raise the level of the ciphertext to $l$, and perform bootstrapping by consuming $l-\delta$ levels using the bootstrapping moduli $q_{\delta+1}, \cdots, q_{L}$.
Let $l_\mathrm{max}$ be the maximum depth of the approximate polynomial for the activation function among all layers in a ResNet model, and let $l_\mathrm{conv}$ be the depth of the convolution operation. Then we have $l_\mathrm{conv}+l_\mathrm{max} = \delta$. If we want to evaluate a layer with approximate polynomial with the depth $l<l_\mathrm{max}$ with this moduli-chain, then the bootstrapping operation is performed in the level using the bootstrapping moduli $q_{\delta+1}, \cdots, q_{L}$, and we drop the $l_\mathrm{max}-l$ levels before evaluating activation function and the convolution. Then, the runtime of the bootstrapping is the same regardless of the depth of the activation function.

Instead of using a single moduli-chain, we propose multiple moduli-chains for each depth of the activation function. If a layer has an approximate polynomial for activation function with the depth $l<l_\mathrm{max}$, the moduli $q_0, q_1, \cdots, q_{l_\mathrm{conv}+l}, q_{\delta + 1}, \cdots, q_L$ are enough for evaluation moduli in the moduli chain for the layer. Then, the bootstrapping operation of this case does not have to operate for moduli $q_{l_\mathrm{conv}+l}, \cdots, q_{\delta}$. It reduces the runtime of the bootstrapping. We can choose which moduli-chain is used for the next layer when the ciphertext level is zero and should be raised with the modulus for the next bootstrapping. Since the coefficients of the secret key are from only $\{-1, 0, 1\}$, the secret key is independent of the moduli, and we can use the several moduli-chains sequentially for one ciphertext encrypted with one secret key.

\textbf{Simulation results.} We validate the amount of time saved through simulation when removing unused moduli during the bootstrapping process in the proposed operations. 
Fig. \ref{fig:moduli} shows the difference of runtime for single bootstrapping and the proposed approximate polynomial evaluation, $T_i(d_i)$ in (\ref{eq:opt_prob}).
Considering the largest degree that consumes the level $\delta$ is $2^\delta - 1$, we perform the approximate polynomials for the degree $d_i=2^\delta - 1$ with $\delta=2,3,\cdots,8$.
Also we set $l_{\text{conv}}=2$, $l_{\text{max}}=8$, and $l_{\text{Boot}}=14$.
For each $\delta$, we compare the runtime for bootstrapping and the following polynomial evaluation consuming depth of $\delta$. 
The result of our simulation shows that removing unused moduli boosts the time of bootstrapping and evaluating approximate polynomials significantly.

\section{Additional Experiments}
\label{app:exp}
To verify that the error in the small-probability region, mentioned in Section \ref{subsec:scaledapp}, significantly impacts performance degradation, we design an experiment to determine whether the small-probability region indeed amplifies much error. We try to infer data by approximating the exact ReLU activation in the \textit{specific region} $R$. That is, $u_R(x) := P_{\phi}[d;f](x)$ when $x \in R$, and $\text{ReLU}(x)$ otherwise. It is reasonable to conclude that the approximate region $R$ affects the performance more than other regions if $u_R(x)$ results in a large classification loss.

In our pre-trained ResNet-20, the first ReLU function has an input distribution $\mathcal{N}(\mu,\sigma^2)$ where $\mu=0.329$ and $\sigma=1.948$. Also, we observe that the input values of the first ReLU for the entire training data lie on the interval $[-11.304,11.820]$. With these data, we evaluate the value of the loss function when the first ReLU function is replaced by $u_R(x)$ for $R=[-11,-10],[-10,-9],\cdots,[10,11]$. We conduct experiments using cross-entropy loss as the classification loss.

\begin{figure}[!t]
  \begin{center}   \includegraphics[width=0.8\textwidth]{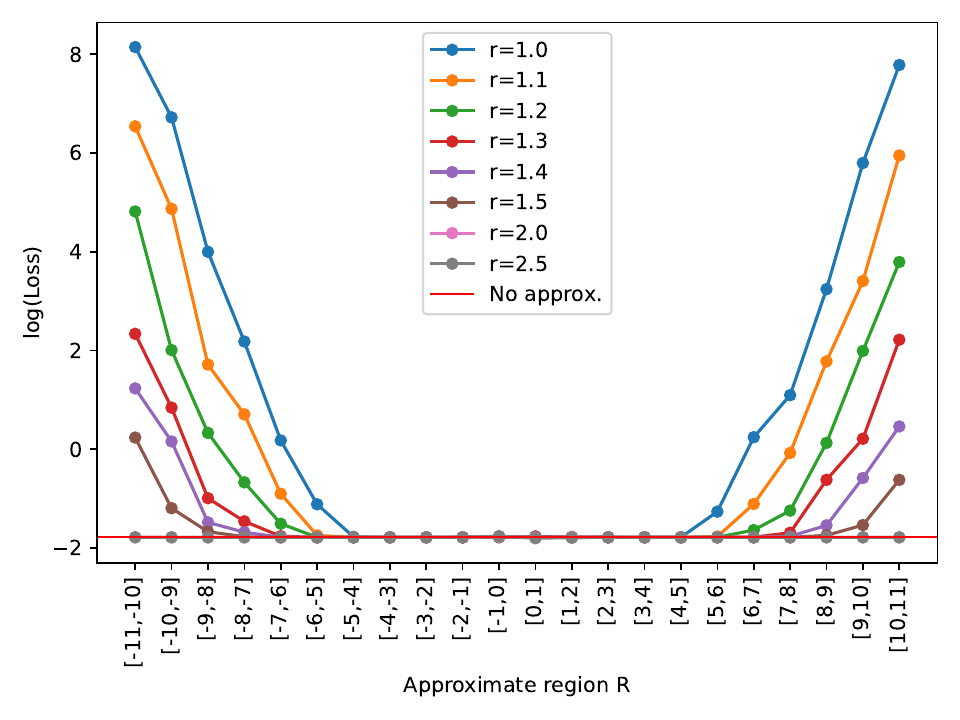}
  \end{center}
  \vspace{-20pt}
  \caption{Graph of the logarithm of the loss for the approximation region $R$ according to changes in the value of $r$.}
  \label{fig:apploss}
\end{figure}

Figure \ref{fig:apploss} shows the logarithm of the loss for each approximation region obtained through our experiments. In the case where $r=1$ on the graph, that is, when the scaled distribution-aware approximation in Section \ref{sec:OLA} is not applied, the loss value diverges significantly, exceeding $10^8$, when approximating the ReLU function in an approximation region far from $\mu=0.329$. On the other hand, when the approximation region is close to $\mu$, the loss value is almost the same as when the ReLU function is not approximated at all. Based on these experimental results, we can conclude that regions far from $\mu$ are much more sensitive to classification.

If we conduct experiments by appropriately varying the value of $r$ and applying the approximate polynomial in a specific region $x\in R$ as $u_{R,r}(x):=P_{\Phi^{[r]}}[d;f](x)$, we can observe that there is little change in the loss value regardless of whether the approximation region is close to or far from $\mu$. This indicates that applying the scaled distribution-aware approximation is valid for polynomial approximation techniques that are robust in low-probability regions.

\bibliographystyle{elsarticle-num} 
\bibliography{bibi_mark}

\begin{thebibliography}{10}
\expandafter\ifx\csname url\endcsname\relax
  \def\url#1{\texttt{#1}}\fi
\expandafter\ifx\csname urlprefix\endcsname\relax\def\urlprefix{URL }\fi
\expandafter\ifx\csname href\endcsname\relax
  \def\href#1#2{#2} \def\path#1{#1}\fi

\bibitem{gilad2016cryptonets}
R.~Gilad-Bachrach, N.~Dowlin, K.~Laine, K.~Lauter, M.~Naehrig, J.~Wernsing, Cryptonets: Applying neural networks to encrypted data with high throughput and accuracy, in: Proc. Int. Conf. Machine Learning, PMLR, 2016, pp. 201--210.

\bibitem{ngraph}
F.~Boemer, Y.~Lao, R.~Cammarota, C.~Wierzynski, {nGraph-HE}: A graph compiler for deep learning on homomorphically encrypted data, in: Proceedings of the 16th ACM International Conference on Computing Frontiers, 2019, pp. 3--13.

\bibitem{HCNN_GPU}
A.~Al~Badawi, J.~Chao, J.~Lin, C.~F. Mun, J.~Jie~Sim, B.~H.~M. Tan, X.~Nan, K.~Mi~Mi~Aung, V.~Ramaseshan~Chandrasekhar, Towards the {AlexNet} moment for homomorphic encryption: {HCNN}, the first homomorphic {CNN} on encrypted data with {GPUs}, IEEE Transactions on Emerging Topics in Computing (2020) 1330--1343.

\bibitem{strided}
E.~Lee, J.-W. Lee, J.~Lee, Y.-S. Kim, Y.~Kim, J.-S. No, W.~Choi, Low-complexity deep convolutional neural networks on fully homomorphic encryption using multiplexed convolutions, in: Proc. Int. Conf. Machine Learning (ICML), PMLR, 2022, pp. 12403--12422.

\bibitem{lee2023precise}
J.~Lee, E.~Lee, J.-W. Lee, Y.~Kim, Y.-S. Kim, J.-S. No, Precise approximation of convolutional neural networks for homomorphically encrypted data, IEEE Access (2023) 62062--62076.

\bibitem{kim2023optimized}
D.~Kim, C.~Guyot, Optimized privacy-preserving cnn inference with fully homomorphic encryption, IEEE Transactions on Information Forensics and Security 18 (2023) 2175--2187.

\bibitem{pmlr-v202-lee23m}
S.~Lee, G.~Lee, J.~W. Kim, J.~Shin, M.-K. Lee, {HETAL}: Efficient privacy-preserving transfer learning with homomorphic encryption, in: Proc. Int. Conf. Machine Learning, Vol. 202 of Proceedings of Machine Learning Research, PMLR, 2023, pp. 19010--19035.

\bibitem{lee2022privacy}
J.-W. Lee, H.~Kang, Y.~Lee, W.~Choi, J.~Eom, M.~Deryabin, E.~Lee, J.~Lee, D.~Yoo, Y.-S. Kim, et~al., Privacy-preserving machine learning with fully homomorphic encryption for deep neural network, IEEE Access 10 (2022) 30039--30054.

\bibitem{FasterCryptoNets}
E.~Chou, J.~Beal, D.~Levy, S.~Yeung, A.~Haque, L.~Fei-Fei, {Faster Cryptonets}: Leveraging sparsity for real-world encrypted inference, arXiv preprint arXiv:1811.09953 (2018).

\bibitem{ali2020polynomial}
R.~E. Ali, J.~So, A.~S. Avestimehr, On polynomial approximations for privacy-preserving and verifiable relu networks, arXiv preprint arXiv:2011.05530 (2020).

\bibitem{leeminimax}
E.~Lee, J.-W. Lee, J.-S. No, Y.-S. Kim, Minimax approximation of sign function by composite polynomial for homomorphic comparison, IEEE Transactions on Dependable and Secure Computing 19~(6) (2021) 3711--3727.

\bibitem{ju2024neujeans}
J.~H. Ju, J.~Park, J.~Kim, M.~Kang, D.~Kim, J.~H. Cheon, J.~H. Ahn, Neujeans: Private neural network inference with joint optimization of convolution and fhe bootstrapping, in: Proceedings of the 2024 on ACM SIGSAC Conference on Computer and Communications Security, 2024, pp. 4361--4375.

\bibitem{zhang2024secure}
J.~Zhang, X.~Yang, L.~He, K.~Chen, W.-j. Lu, Y.~Wang, X.~Hou, J.~Liu, K.~Ren, X.~Yang, Secure transformer inference made non-interactive, Cryptology ePrint Archive (2024).

\bibitem{moon2024thor}
J.~Moon, D.~Yoo, X.~Jiang, M.~Kim, Thor: Secure transformer inference with homomorphic encryption, Cryptology ePrint Archive (2024).

\bibitem{park2024powerformer}
D.~Park, E.~Lee, J.-W. Lee, Powerformer: Efficient and high-accuracy privacy-preserving language model with homomorphic encryption, Cryptology ePrint Archive (2024).

\bibitem{ran2023spencnn}
R.~Ran, X.~Luo, W.~Wang, T.~Liu, G.~Quan, X.~Xu, C.~Ding, W.~Wen, Spencnn: orchestrating encoding and sparsity for fast homomorphically encrypted neural network inference, in: International Conference on Machine Learning, PMLR, 2023, pp. 28718--28728.

\bibitem{kim2023hyphen}
D.~Kim, J.~Park, J.~Kim, S.~Kim, J.~H. Ahn, Hyphen: A hybrid packing method and its optimizations for homomorphic encryption-based neural networks, IEEE Access (2023) 3024--3038.

\bibitem{luo2024secformer}
J.~Luo, Y.~Zhang, Z.~Zhang, J.~Zhang, X.~Mu, H.~Wang, Y.~Yu, Z.~Xu, Secformer: fast and accurate privacy-preserving inference for transformer models via smpc, in: Findings of the Association for Computational Linguistics ACL 2024, 2024, pp. 13333--13348.

\bibitem{li2022mpcformer}
D.~Li, R.~Shao, H.~Wang, H.~Guo, E.~P. Xing, H.~Zhang, Mpcformer: fast, performant and private transformer inference with mpc, arXiv preprint arXiv:2211.01452 (2022).

\bibitem{maeng2024accelerating}
K.~Maeng, G.~E. Suh, Accelerating relu for mpc-based private inference with a communication-efficient sign estimation, Proceedings of Machine Learning and Systems 6 (2024) 128--147.

\bibitem{kei2025shaft}
A.~Y. Kei, S.~S. Chow, Shaft: Secure, handy, accurate, and fast transformer inference, in: Network and Distributed System Security Symposium, NDSS, Vol. 2025, 2025.

\bibitem{xia2025cryptpeft}
S.~Xia, W.~Wang, Z.~Wang, Y.~Zhang, Y.~Jin, D.~Meng, R.~Hou, Cryptpeft: Efficient and private neural network inference via parameter-efficient fine-tuning, arXiv preprint arXiv:2508.12264 (2025).

\bibitem{yuan2024md}
B.~Yuan, S.~Yang, Y.~Zhang, N.~Ding, D.~Gu, S.-F. Sun, $\{$MD-ML$\}$: Super fast $\{$Privacy-Preserving$\}$ machine learning for malicious security with a dishonest majority, in: 33rd USENIX Security Symposium (USENIX Security 24), 2024, pp. 2227--2244.

\bibitem{cheng2025mosformer}
K.~Cheng, Y.~Xia, A.~Song, J.~Fu, W.~Qu, Y.~Shen, J.~Zhang, Mosformer: Maliciously secure three-party inference framework for large transformers, Cryptology ePrint Archive (2025).

\bibitem{park2022aespa}
J.~Park, M.~J. Kim, W.~Jung, J.~H. Ahn, Aespa: Accuracy preserving low-degree polynomial activation for fast private inference, arXiv preprint arXiv:2201.06699 (2022).

\bibitem{huang2022cheetah}
Z.~Huang, W.-j. Lu, C.~Hong, J.~Ding, Cheetah: Lean and fast secure $\{$Two-Party$\}$ deep neural network inference, in: 31st USENIX Security Symposium (USENIX Security 22), 2022, pp. 809--826.

\bibitem{xu2024Privcirnet}
T.~Xu, L.~Wu, R.~Wang, M.~Li, Privcirnet: Efficient private inference via block circulant transformation, Advances in Neural Information Processing Systems 37 (2024) 111802--111831.

\bibitem{pang2024bolt}
Q.~Pang, J.~Zhu, H.~M{\"o}llering, W.~Zheng, T.~Schneider, Bolt: Privacy-preserving, accurate and efficient inference for transformers, in: 2024 IEEE Symposium on Security and Privacy (SP), IEEE, 2024, pp. 4753--4771.

\bibitem{hao2022iron}
M.~Hao, H.~Li, H.~Chen, P.~Xing, G.~Xu, T.~Zhang, Iron: Private inference on transformers, Advances in neural information processing systems 35 (2022) 15718--15731.

\bibitem{xu2025breaking}
T.~Xu, W.-j. Lu, J.~Yu, Y.~Chen, C.~Lin, R.~Wang, M.~Li, Breaking the layer barrier: Remodeling private transformer inference with hybrid $\{$CKKS$\}$ and $\{$MPC$\}$, in: 34th USENIX Security Symposium (USENIX Security 25), 2025, pp. 2653--2672.

\bibitem{mishra2020delphi}
P.~Mishra, R.~Lehmkuhl, A.~Srinivasan, W.~Zheng, R.~A. Popa, Delphi: A cryptographic inference system for neural networks, in: Proceedings of the 2020 Workshop on Privacy-Preserving Machine Learning in Practice, 2020, pp. 27--30.

\bibitem{cho2024fast}
W.~Cho, G.~Hanrot, T.~Kim, M.~Park, D.~Stehl{\'e}, Fast and accurate homomorphic softmax evaluation, in: Proceedings of the 2024 on ACM SIGSAC Conference on Computer and Communications Security, 2024, pp. 4391--4404.

\bibitem{meftah2021doren}
S.~Meftah, B.~H.~M. Tan, C.~F. Mun, K.~M.~M. Aung, B.~Veeravalli, V.~Chandrasekhar, Doren: toward efficient deep convolutional neural networks with fully homomorphic encryption, IEEE Transactions on Information Forensics and Security 16 (2021) 3740--3752.

\bibitem{lou2019she}
Q.~Lou, L.~Jiang, She: A fast and accurate deep neural network for encrypted data, Advances in neural information processing systems 32 (2019).

\bibitem{iccv}
H.~Peng, S.~Huang, T.~Zhou, Y.~Luo, C.~Wang, Z.~Wang, J.~Zhao, X.~Xie, A.~Li, T.~Geng, et~al., Autorep: Automatic relu replacement for fast private network inference, in: Proceedings of the IEEE/CVF International Conference on Computer Vision, 2023, pp. 5178--5188.

\bibitem{lattigo}
Lattigo, Online: \url{https://github.com/tuneinsight/lattigo/tree/v3.0.5}, ePFL-LDS, Tune Insight SA (Apr. 2022).

\bibitem{krizhevsky2009learning}
A.~Krizhevsky, G.~Hinton, et~al., Learning multiple layers of features from tiny images, CiteSeerX Technical Report, University of Toronto (2009).

\bibitem{russakovsky2015imagenet}
O.~Russakovsky, J.~Deng, H.~Su, J.~Krause, S.~Satheesh, S.~Ma, Z.~Huang, A.~Karpathy, A.~Khosla, M.~Bernstein, et~al., Imagenet large scale visual recognition challenge, International Journal of Computer Vision 115~(3) (2015) 211--252.

\bibitem{he2016deep}
K.~He, X.~Zhang, S.~Ren, J.~Sun, Deep residual learning for image recognition, in: Proceedings of the IEEE Conference on Computer Vision and Pattern Recognition, 2016, pp. 770--778.

\bibitem{liu2022convnet}
Z.~Liu, H.~Mao, C.-Y. Wu, C.~Feichtenhofer, T.~Darrell, S.~Xie, A convnet for the 2020s, in: Proceedings of the IEEE/CVF conference on computer vision and pattern recognition, 2022, pp. 11976--11986.

\bibitem{hendrycks2016gaussian}
D.~Hendrycks, K.~Gimpel, Gaussian error linear units (gelus), arXiv preprint arXiv:1606.08415 (2016).

\bibitem{RNS-CKKS}
J.~H. Cheon, K.~Han, A.~Kim, M.~Kim, Y.~Song, A full rns variant of approximate homomorphic encryption, in: International Conference on Selected Areas in Cryptography, Springer, 2018, pp. 347--368.

\bibitem{jha2021deepreduce}
N.~K. Jha, Z.~Ghodsi, S.~Garg, B.~Reagen, Deepreduce: Relu reduction for fast private inference, in: International Conference on Machine Learning, PMLR, 2021, pp. 4839--4849.

\bibitem{cheon2024batch}
J.~H. Cheon, M.~Kang, T.~Kim, J.~Jung, Y.~Yeo, Batch inference on deep convolutional neural networks with fully homomorphic encryption using channel-by-channel convolutions, IEEE Transactions on Dependable and Secure Computing (2024) 1--12.

\bibitem{vaswani2017attention}
A.~Vaswani, N.~Shazzer, N.~Parmar, J.~Uszkoreit, L.~Jones, A.~N. Gomez, L.~Kaiser, I.~Polosukhin, Attention is all you need, Advances in Neural Information Processing Systems (2017).

\bibitem{bossuat2021efficient}
J.-P. Bossuat, C.~Mouchet, J.~Troncoso-Pastoriza, J.-P. Hubaux, Efficient bootstrapping for approximate homomorphic encryption with non-sparse keys, in: Annual International Conference on the Theory and Applications of Cryptographic Techniques, Springer, 2021, pp. 587--617.

\end{thebibliography}


%\begin{thebibliography}{00}

%% For numbered reference style

%\end{thebibliography}
\end{document}